\documentclass[a4paper,11pt]{article}
\pdfoutput=1 
\usepackage{jheppub} 
\usepackage{mciteplus}
\usepackage{dirtytalk}

\usepackage{tikz,xcolor,hyperref}

\usepackage{amsmath, amssymb, amsthm, graphicx, epsfig, fancyhdr,epsfig, slashed, mathrsfs}

\usepackage{tikzsymbols}
\usepackage{natbib}
\usepackage{float}

\usepackage{mathtools} 
\newcommand{\ba}{\begin{eqnarray}}
\newcommand{\ea}{\end{eqnarray}}
\newcommand{\be}{\begin{equation}}
\newcommand{\ee}{\end{equation}}
\newcommand{\bi}{\begin{itemize}}
\newcommand{\ei}{\end{itemize}}

\newcommand{\la}{\lambda}

\newcommand{\sa}{\sigma}

\newcommand{\Fc}{\mathcal{F}}
\newcommand{\Gc}{\mathcal{G}}

\newcommand{\Lc}{\mathcal{L}}
\newcommand{\Mc}{\mathcal{M}}

\newcommand{\pd}{\partial}

\newcommand{\const}{\text{const}}

\title{Non-perturbative quantum gravity denounces singular Black Holes}

\author[a,b]{Alexey S. Koshelev,}

\author[c,d,e]{Anna Tokareva}

\affiliation[a~]{
School of Physical Science and Technology, ShanghaiTech University, 201210 Shanghai, China
}
\affiliation[b~]{
	Departamento de F\'isica, Centro de Matem\'atica e Aplica\c{c}oes (CMA-UBI),
	Universidade da Beira Interior, 6200 Covilh\~a, Portugal }
\affiliation[c~]{School of Fundamental Physics and Mathematical Sciences, Hangzhou Institute for Advanced Study, UCAS, Hangzhou 310024, China}
\affiliation[d~]{International Centre for Theoretical Physics Asia-Pacific, Beijing/Hangzhou, China}
\affiliation[e~]{Theoretical Physics, Blackett Laboratory, Imperial College London, SW7 2AZ London, U.K.}

\emailAdd{askoshelev@shanghaitech.edu.cn}
\emailAdd{tokareva@ucas.ac.cn}

\abstract{

Although General Relativity predicts the presence of a singularity inside of a Black Hole, it is not a complete theory of gravity. A real structure of a Black Hole interior near an expected singularity depends on the UV completion of gravity. In this paper we establish that the question whether singular spherically symmetric solutions are absent is governed by the functional form of a non-perturbative graviton propagator. We explicitly show in a framework of a ghost-free infinite derivative gravity that for the graviton propagator of an exponential form favored by the unitarity a singularity is not possible unless an unphysical situation when the total mass of the Black Hole is infinite is considered.
}

\begin{document}

\maketitle

\section{Introduction}

Einstein's General Relativity (GR) \cite{Wald:1984rg} perfectly captures IR gravitational physics but fails at high energies, or equivalently at small distances. However, its low energy predictions are well studied and heavily scrutinized by many very accurate measurements. Schwarzschild metric \cite{Schwarzschild:1916uq} is a notorious solution to GR equations of motion describing a compact centrally symmetric massive object in an empty space. This metric greatly explains the Mercury perihelion displacement -- the problem which was one of the original motivation of Einstein to construct a new gravity theory -- but also manifests the problem of singularity. It has become the first solution for a Black Hole (BH), the Schwarzschild BH (SBH), which is the simplest static BH solution without a charge. The center of the SBH appears to be a singularity of the space-time. Before elaborating more on the notion of singularity we note that this is obviously related to the failure of GR to describe high energies and there is a natural expectation that a UV complete gravity should avoid singularities, including the BH ones.

Surely, term \textit{singularity} prompts for more explanation. A recent paper by Kerr \cite{Kerr:2023rpn} provides a very nice and exciting discussion on the topic. We adopt a good starting proposition stated in that paper and say that a singularity is a point or a region in the space-time where metric or curvature are unbounded, or non-differentiable a required number of times. However, not every singularity is physical and some can be removed by a better choice of coordinates like for example, Eddington-Finkelstein \cite{Eddington:1924pmh,Finkelstein:1958zz} coordinates for the SBH which demonstrate that there is no singularity at the horizon. A much more problematic singularity is the one which cannot be removed by changing coordinates. It is a real, or physical singularity which appears in some geometrical scalar invariants. Scalar curvature $R$ is the simplest scalar invariant but it is not the easiest one to use in the case of the SBH at least as its value cannot be straightforwardly determined at the origin. Naively it is zero everywhere but there is some entity curving spacetime anyway. To make sense of this one can either puncture the manifold at zero \cite{Hawking:1973uf} or consider the SBH in a distributional sense \cite{Balasin:1993fn,Steinbauer:2006qi}. This issue about $R$ is an integral part of our discussion and we will return to it in due course. A more apparent invariant, the Kretschmann scalar $K=R_{\mu\nu\alpha\beta}R^{\mu\nu\alpha\beta}$, behaves as $K\sim 1/r^6$ for the SBH in 4 dimensions and clearly shows that the origin is a singularity. In what follows we focus on the absence (or presence) of physical singularities in BH solutions and use the term \textit{singularity} for physical singularities unless specified otherwise.

Here we emphasize that on the way of constructing new solutions in GR starting from Einstein equations $G_{\mu\nu}=8\pi GT_{\mu\nu}$, where $G$ is the Newtonian constant, one can try to find a geometry for a given matter content or alternatively compute a matter content which corresponds to a suggested geometry. While the former is an almost unfeasible task in general the latter can be (almost always) readily implemented by substituting a given metric in the LHS (i.e. geometrical part) of equations. This is also a usual approach for judging whether new solutions correspond to a physically justifiable matter on the RHS (i.e. matter part) of equations. For example, whether it is indeed a compact mass distribution, whether energy conditions are satisfied, etc. Various attempts to find regular BH solutions \cite{Bardeen:1968qqq,Frolov:1988vj,Dymnikova:1992ux,Ayon-Beato:1998hmi,Bonanno:2000ep,Bronnikov:2000vy,Hayward:2005gi}  and to improve the SBH or other singular solutions in the framework of GR such that a singularity disappears lead to a necessity of some non-point-like and moreover non-compact mass distribution on the RHS of Einstein equations or specific often non-linear sources (see \cite{Ansoldi:2008jw} for a review). It is realistic however to come up with a compact object solution by gluing a Schwarzschild solution beyond some radius and a modified solutions inside some radius \cite{Janis:1965tx,10.1063/1.525325,Erbin:2016lzq} and this procedure is possible because equations contain two derivatives and thus only metric functions and their first derivatives must be glued, like in quantum mechanics.

Turning to the idea of a singularity resolution in a UV complete gravity theory \cite{Buoninfante:2018rlq,dePaulaNetto:2023vtg,Kolar:2023gqi} we start by noting that high energy properties can be readily improved by going to a quadratic curvature gravity. This approach initially formulated in \cite{Stelle:1976gc} unfortunately suffers from the presence of a ghost. One can go further and consider a gravity theory with infinite derivative operators often named analytic infinite derivative gravity \cite{Kuzmin:1989sp,Tomboulis:1997gg,Biswas:2011ar,Koshelev:2017ebj}. It is important to jump to an infinite number of derivatives because any finite number of derivatives will result in ghosts, as we perfectly know for more than 170 years thanks to Ostrogradski \cite{Ostrogradsky:1850fid}. An approach of asymptotic safety in gravity \cite{Percacci:2017fkn,Reuter:2019byg} also leads to generalizations of gravity containing functions of the d'Alembertian operator \cite{Knorr:2021iwv, Fehre:2021eob, Platania:2022gtt} in the non-perturbative effective action. Numerous considerations in the effective field theory framework \cite{Donoghue:1995cz,Burgess:2003jk, Cardoso:2018ptl, Serra:2022pzl, deRham:2021bll, Cano:2019ore, Melville:2024zjq} equally suggest the inevitable appearance and importance of higher derivative operators in gravity modifications.

We consider the presence of infinite derivatives as a necessary condition to build a UV complete quantum gravity. Before going on with actual computations regarding BH-s in such a setup it is important to stress that the presence of infinite derivatives results in new obstacles. First, on contrary to a finite derivative situation just a would-be routine of substituting a given metric in equations of motion is already a non-trivial and generically impossible operation. Even though it looks strange, one cannot always obtain a resulting RHS of generalized Einstein equations in this case even when a metric is given explicitly. Second, infinite derivative operators demand to have infinitely differentiable functions. This almost ultimately scraps a possibility to glue different solutions in different regions because now we must glue all and not just two derivatives. Third, the problem of being singular or not is now much more involved. As mentioned above, in GR usually the Kretschmann invariant is used as an indicator of the regularity of a space-time. We will demonstrate below that one can easily come to a situation when $K=R_{\mu\nu\alpha\beta}R^{\mu\nu\alpha\beta}$ is regular and $K_1=R_{\mu\nu\alpha\beta}\Box R^{\mu\nu\alpha\beta}$ is regular as well but $K_2=R_{\mu\nu\alpha\beta}\Box^2R^{\mu\nu\alpha\beta}$ is not for some known regular BH candidates in GR. Here $\Box$ is the covariant d'Alembertian and $K_2$ is obviously a scalar. An obvious explanation of why quantities like $K_2$ were not tested in GR is just because terms like this do not show up in any computation of physically important quantities. It is not the case when there are infinite derivatives though as now we encounter some endless zoo of terms.

In fact, already with a finite number of derivatives seeking for BH solutions is a very difficult task which involves a tough equation analysis. In the case of the quadratic gravity, this was done for example in \cite{Nelson:2010ig,Lu:2015cqa,Lu:2015psa,Lu:2017kzi,Holdom:2016nek,Cai:2015fia,Saueressig:2021wam}. BH-s in asymptotically safe gravity are nicely reviewed in \cite{Koch:2014cqa} and more recently in \cite{Platania:2023srt} where references therein point to a number of important studies. BH-s in higher and infinite derivative gravity models were considered in, for instance, \cite{Calcagni:2018pro,Burzilla:2023xdd,dePaulaNetto:2023vtg}. In these studies gravity modifications do not contain a 4-rank tensor simplifying an analysis. An example study showing that the SBH solution cannot be in a higher derivative (with more than four derivatives) gravity with a Weyl tensor can be found in \cite{Koshelev:2018hpt}. Properties of some non-locally modified BH-s were considered in \cite{Frolov:2015bia,Frolov:2023jvi}. In the majority of written papers, linearized equations were considered which seem to be way incompatible with the strong gravity regime near a singularity. Also often the main focus of other papers was a construction of regular solutions rather than the study of their general properties.

In the present paper, we challenge a general question of whether singular static centrally symmetric solutions for the metric are ever possible in an infinite derivative gravity considering full non-linear infinite derivative equations. We assume that a singular solution can be considered to be a valid solution only if it can be obtained as a limiting case of regular solutions and in this limit corresponds to a physically viable matter distribution. This assumption is inspired by the idea that a real astrophysical BH should be a result of a matter collapse that starts from a perfectly regular configuration.

$~$

A summary of our analysis strategy is as follows.

First, we describe a quadratic in curvature infinite derivative gravity model which is under consideration throughout this paper. One of the main specific ingredient of such models are higher and infinite derivative operators (or form-factors) represented as series of the d'Alembertian operator. Then we introduce a BH metric of the form
\begin{equation*}
    ds^2=-A(r)dt^2+A(r)^{-1}dr^2+r^2 d\Omega^2
\end{equation*}
where we choose the only yet undetermined function $A(r)$ for simplicity and present some general comments about BH-s in an infinite derivative gravity.
This constitutes Section~2.

Second, we specialize to the SBH for which most of the analysis of this paper is done and which is characterized by
\begin{equation*}
    A(r)=1-\frac{2GM}r
\end{equation*}
where $G$ is the Newtonian constant and $M$ is the BH mass. We explain why nonlinearities present difficulties and also pay a special attention to the fact that infinite derivatives can smear even a point-support function.
This forms Section~3.

Third, we introduce a regularization of the SBH described by a parameter $\alpha$ such that $\alpha\to0$ corresponds to a restoration of the original SBH solution. It is already a special task to find such a regularization that all higher derivative terms in full non-linear equations become regular. We come up with a family of exponential regularizations of the form
\begin{equation*}
    A=1-\frac{2GM}r\exp(-\alpha r^{-p}),\quad p>0
\end{equation*}
Even though we cannot compute the RHS of equations completely as we cannot manage accounting an infinite number of terms, we demonstrate that every term in the equations of motion is regular. This comprises Section~4.

Fourth, we turn to a physically important quantity of the full energy which is simply a mass of our centrally symmetric object. It is an integral of the energy density over the 3-dimensional space. This is equivalently given by integrating the LHS of the generalized Einstein equations. Doing so before removing the regularization we observe that terms in EOM-s coming from the original Einstein-Hilbert action become upon integration a finite constant which is precisely the mass parameter $M$ in the SBH metric and also this term does not depend on $\alpha$; terms in EOM-s coming from local quadratic in curvature terms in the action do not contribute to the total energy at all; and other terms in EOM-s coming from six- (and higher)-derivative terms in the action produce upon integration results singular in $\alpha$. Corresponding computations are presented in Section~5.

Fifth, it is obvious at this stage that the SBH fits nicely in GR and in the local quadratic gravity, while it is not a physically viable solution in the six-order gravity and beyond but a finite order gravity. However, it is an absolutely unclear and non-trivial question whether in the case of an infinite derivative gravity, an infinite series of terms  with each term singular in the limit $\alpha\to 0$ can be summed up to zero or a finite answer. The resulting series which determines a BH mass has a form $\sum\limits_{i,n}\hat {f_{i}}_n {v_{i}}_n$ where $\hat f$-coefficients are determined by the higher derivative form-factors in the action and $v$-coefficients result from the integration over the space (the meaning of indexes will become clear below in the main text). One thus should find conditions on the series expansion of operators in the gravity model which will result in converging series for a BH mass. A comprehensive study of this involved question using thoroughly the complex analysis is done in Section~6.

We then collect our findings in the conclusion Section discussing in particular the regularization independence of our approach and outlining future directions of study.

\section{Analytic infinite derivative gravity model and Black Holes}

\subsection{Model and Equations of Motion}

We consider a Lagrangian with infinite derivative operators (or form-factors) in a quadratic in curvature action which reads as follows
\begin{equation}
S = \int d^4x\ \sqrt{-g}\left( \frac{M_P^2}2 R 
+\frac\lambda2\left[R
\Fc_1(\Box)R+L_{\mu\nu}
\Fc_2(\Box)L^{\mu\nu}+ W_{\mu\nu\la\sa}
\Fc_{4}(\Box)W^{\mu\nu\la\sa}\right] +\dots
\right)
\label{properaction}
\end{equation}
Here $M_P$ is the Planck mass related to the Newtonian constant $G$ as $M_P^{-2}=8\pi G$, $\lambda$ is a dimensionless parameter that can be used to track the GR limit, 
$  L_{\mu\nu}=R_{\mu\nu}-\frac14g_{\mu\nu}R$ is the traceless part of the Ricci tensor, $W_{\mu\alpha\nu\beta}$ is the Weyl tensor and operators $\Fc_i(\Box)$ are analytic at zero functions represented as
\begin{equation*}
  \Fc_i=\sum_{n=0}^\infty {f_i}_n(\Box/\Mc^2)^n
\end{equation*}
$\Mc$ is the scale for higher derivative modifications and ${f_i}_n$ are constants. In the majority of the computations below $\Mc$ is put to unity. Also in this paper we consider only the case of four dimensions.

The above action is important because it is the most general (and even still redundant) action containing all the possible terms that contribute to a propagator around Minkowski space-time \cite{Biswas:2016egy}. If one considers only a propagator in Minkowski space-time any of the form-factors, say $\Fc_2$, can be proven to be redundant and thus put to be zero in four dimensions.\footnote{If one adds a cosmological term the same statement about the generality of this action is valid around an (anti-)de Sitter space. For either Minkowski or (anti-)de Sitter in $D>4$ if, say, $\Fc_{1,4}$ are kept general one needs to keep also $R_{\mu\nu}^2$ local term with a constant coefficient for generality.} Dots reflect terms which either do not contribute to a propagator or give a contribution equivalent to already written terms upon rescaling of the corresponding couplings. These terms if included will change higher interaction vertices around Minkowski space-time.

While higher derivatives produce ghosts in general, an infinite number of derivatives can evade this problem.
The requirement for the theory to be ghost-free means that the propagator must have no extra poles except the standard one. This can be satisfied only if the propagator has a form (around Minkowski space-time)
\begin{equation}
    \Pi(k^2)=\frac{e^{2\sigma(k^2)}}{k^2},
    \label{prop}
\end{equation}
where function $\sigma(k^2)$ is an entire function. 
Obviously, higher derivatives can improve the UV behavior and thus one can hope for renormalizability and unitarity of this model at once. These considerations lead to a claim that the above action with $\Fc_2=0$ and without extra terms can be a Quantum Gravity candidate which also embeds nicely and exactly a Starobinsky inflationary solution.
{Furthermore, if one requires that the only degree of freedom is the massless graviton upon resolving Hamiltonian constraints, a relation
\begin{equation}
\label{Ucond}
    \Fc_4(\square)=-3\Fc_1(\square)
\end{equation}
should be satisfied. To have an additional scalar, say inflaton, the relation is more complicated, see \cite{Koshelev:2022wqj}.}
On the other token, Quantum Gravity considerations from the point of view of asymptotic safety also lead to the same infinite derivative action form as long as a propagator is concerned.

To say that the above model is credible one needs to have Black Holes as solutions in this model. Obviously, those terms that are hidden in dots in principle may have some contribution for a geometry that is not maximally symmetric. Thus in principle, these other terms should be studied but for now we keep their consideration for future papers. However, since none of the form-factors can be consistently dropped for a BH background, we do not disregard the term containing $\Fc_2$ and it may lead to special configurations as will be shown later.

Equations of motion for action (\ref{properaction}) are long and involved and we will not use them all. They were derived in \cite{Biswas:2013cha} and for the sake of completeness we put them in Appendix~\ref{appeom} of this paper. For our purposes, we will concentrate mostly on the trace equation which reads:
\begin{equation}
\begin{split}
&M_P^2 R
-6\lambda\Box\Fc_1(\Box) R-\lambda({\Lc_1}^\mu_\mu +2\bar\Lc_{1})-\\
-& 2\lambda\nabla_\rho\nabla_\mu\Fc_2(\Box){L}^{\mu\rho}
-\lambda({\Lc_2}^\mu_\mu +2\bar\Lc_{2}) +2\lambda{\Delta_2}^\mu_\mu -\lambda({\Lc_4}+2\bar\Lc_{4})+4\lambda{\Delta_4}^\mu_\mu =-T
\end{split}
\label{EOMtrace}
\end{equation}
Here we denote:
\begin{equation*}
{\Lc_1}^\mu_\nu=\sum_{n=1}
^\infty
{{ f}_1}{}_n\sum_{l=0}^{n-1}\pd^\mu  R^{(l)}  \pd_\nu  R^{(n-l-1)},~
\bar\Lc_1=\sum_{n=1}
^\infty
{ f_1}{}_n\sum_{l=0}^{n-1} R^{(l)}    R^{(n-l)},
\end{equation*}
\begin{equation*}
{\Lc_2}^\mu_\nu=\sum_{n=1}
^\infty
{f_2}_n\sum_{l=0}^{n-1}\nabla^\mu {{L}^{(l)} }{}^\alpha_\beta 
\nabla_\nu  {{L}^{(n-l-1)}}{}^\beta_\alpha
,~\bar\Lc_2=\sum_{n=1}
^\infty
{f_2}_n\sum_{l=0}^{n-1}{L^{(l)}}{}^\alpha_\beta
{L^{(n-l)}}{}^\beta_\alpha,
\end{equation*}
\begin{equation*}
{\Lc_4}^\mu_\nu=\sum_{n=1}
^\infty
{f_4}_n\sum_{l=0}^{n-1}\nabla^\mu {{W}^{(l)}
}{}^\alpha_{\phantom{\alpha}\beta\gamma\delta} 
\nabla_\nu  {{W}^{(n-l-1)}}{}_\alpha^{\phantom{\alpha}\beta\gamma\delta}
,~\bar\Lc_4=\sum_{n=1}
^\infty
{f_4}_n\sum_{l=0}^{n-1}{W^{(l)}}{}^\alpha_{\beta\gamma\delta}
{W^{(n-l)}}{}_\alpha^{\phantom{\alpha}\beta\gamma\delta},
\end{equation*}
\begin{equation*}
{\Delta_2}^\mu_{\nu}=
\sum_{n=1}
^\infty
{f_2}_n\sum_{l=0}^{n-1}\nabla_\beta[ {L^{(l)}}{}^\beta_\gamma
\nabla^\mu{L^{(n-l-1)}}{}^\gamma_\nu-\nabla^\mu
{L^{(l)}}{}^\beta_\gamma {L^{(n-l-1)}}{}^\gamma_\nu]
\end{equation*}
\begin{equation*}
{\Delta_4}^\mu_{\nu}=
\sum_{n=1}
^\infty
{f_4}_n\sum_{l=0}^{n-1}\nabla_\beta[
{{W}^{(l)}}{}^{\beta\rho}_{\phantom{\beta\rho}\gamma\zeta}
\nabla^\mu{{W}^{(n-l-1)}}{}^{\phantom{\nu\rho}\gamma\zeta}_{\nu\rho}
-\nabla^\mu
{{W}^{(l)}}{}^{\beta\rho}_{\phantom{\beta\rho}\gamma\zeta}
{{W}^{(n-l-1)}}{}^{\phantom{\nu\rho}\gamma\zeta}_{\nu\rho}]
\end{equation*}
and use a convention $X^{(n)}\equiv \Box^n X$ for any tensor $X$.
The RHS of the above equation contains the trace $T\equiv T^\mu_\mu$ of an energy-momentum tensor of matter minimally coupled to gravity.

There is no need to explain that solving this equation in general is far beyond our abilities. Even a simple case of $T=0$ is not simple at all. However, we know several examples of exact and non-trivial solutions (as a trivial solution, Minkowski is an obvious background in this model): for example the Starobinsky inflation when a radiation source on the RHS is added; or bouncing solutions if a cosmological term and radiation are added. In these examples solutions are conformally flat obeying a condition
\begin{equation}
\Box R=m^2 R
\label{condition}
\end{equation}
where $m$ is some mass (of an inflaton in the case of inflation).

\subsection{Preliminary comments on Black Holes}

To describe a BH we need to have a horizon manifesting itself as a coordinate singularity in the metric. Further, we should encounter a singularity or an inner horizon and a regular interior so that propositions of Hawking-Penrose theorems \cite{Penrose:1964wq,Hawking:1971vc} are fulfilled.
To find a BH solution we assume a spherically symmetric ansatz for the metric
\begin{equation}
    ds^2=-A(r)dt^2+A(r)^{-1}dr^2+r^2 d\Omega^2
    \label{startmetricagain}
\end{equation}
Here from the very beginning we will consider a simplified configuration such that function $A(r)$ is one and the same in front of $dt^2$ and $dr^2$. This definitely can be relaxed and deserves more study in forthcoming publications. For a BH case the above metric cannot be made conformally flat. Indeed, conformal flatness is equivalent to having a vanishing identically Weyl tensor. Thus, one can consider Weyl tensor components. For the above metric Weyl tensor has a form $W^{\mu\nu}_{\phantom{\mu\nu}\alpha\beta}=w(r)C^{\mu\nu}_{\phantom{\mu\nu}\alpha\beta}$ where
\begin{equation}
    w=r^2A''(r)-2rA'(r)+2A(r)-2
\label{weylfactor}
\end{equation}
and all non-zero components of $C^{\mu\nu}_{\phantom{\mu\nu}\alpha\beta}$ are numeric constants.
Hereafter we use prime to denote a derivative w.r.t. $r$.
To make all the components of Weyl tensor zero we have to solve $w=0$ for $A(r)$ which has a general solution $A(r)=1+c_1r+c_2r^2$ not suitable for our needs. From here we conclude that a possible simplification $\Fc_4=0$ can not be justified unless imposed by hands.

Next one can try to explore a tantalizing possibility to make use of relation (\ref{condition}) or some of its generalization to higher (but finite) powers of d'Alemebrtian and higher rank tensors. While not excluded, finding an exact analytic form of function $A(r)$ which also describes a reasonable geometry seems to be an extremely involved problem without a known solution yet. However a much more important note is in order here. Relation (\ref{condition}) is in fact a trace equation of a local $R^2$ gravity. It means that satisfying this equation one picks up a solution of a local $R^2$ gravity and embeds it in an infinite derivative model. A more general relation with more derivatives will correspond to, say, six-order gravity or eight-, or more. This essentially reduces an infinite derivative model to a finite-derivative one. This may be not compatible as we will see below that there is a huge jump between situations with finite and infinite number of derivatives as long as BH-s are considered.

Let us make several comments here. First, solving what would be the most relevant task, i.e. finding a metric when the matter content is given, is obviously a problem of extreme difficulty definitely beyond the scope of this paper. Second, even if a metric is given explicitly there is no predefined way of substituting it into our equations analytically. Every new order in d'Alembertian will generate new structures. There is no systematic way to proceed unless some recursion appears. However, as commented above, a recursion may degrade the system to a different, higher but finite-order derivative model. Third, despite these odds it appears to be realistic to analyze singular space-times and demonstrate their incompatibility with the presence of higher or infinite number of derivatives. Namely, in the case of finite higher derivatives and in the case of well-justified from the quantization point of view infinite derivative form-factors the corresponding matter content will be a dubious infinite mass BH, not necessarily compact.

\section{Schwarzschild Black Hole, point matter source and infinite derivatives}


From a physics point of view, one naturally expects the Schwarzschild BH (SBH) to be a solution here. At least far away from a massive compact object at distances much larger than $1/\Mc$ our model reduces to GR and thus the SBH solution should be restored. However close to the origin, at distances of order $1/\Mc$ (which are very much expected to be inside of an outer horizon) higher derivatives take over and may bring new effects.

One can trivially repel the SBH solution altogether on the basis of the presence of the Weyl tensor. The SBH is given in our notations by
\begin{equation}
    A(r)=1-\frac{2GM}r
\end{equation}
where $M$ is the BH mass and the Schwarzschild radius is $r_0=2GM$. As shown in the previous Section the corresponding Weyl tensor is not zero and this will prevent it from being a vacuum solution according to the analysis done in \cite{Koshelev:2018hpt}.

It is possible to try harder and to impose a condition $\Fc_4=0$. The model still can be ghost-free in Minkowski space-time but the Starobinsky inflation will not be a solution there. Nevertheless one will be left with equations containing only Ricci scalar and Ricci tensor but not Weyl tensor.

Naively one can use the fact that $R_{\mu\nu}=R=0$ for SBH but one should understand that this is true at any point but the origin. Furthermore, removing a single point like it is often done in studying BH-s in GR \cite{Hawking:1973uf} is not a valid procedure here because infinite derivatives are involved. Infinite derivatives effectively convert our equations to some sort of integral equations and this should be considered as a boundary value problem. In the latter case, all the manifold must be under consideration.

An alternative to removing singular points from the manifold is a consideration of the SBH in a distributional sense \cite{Balasin:1993fn,Steinbauer:2006qi}. This approach adds a point-like mass source of an infinite delta-function-like density at the origin. This directly attributes mass to the physical matter. Then from the trace of Einstein equations, we get
\begin{equation}
    R\propto T \propto M\delta^3(r)
\end{equation}
where $M$ is exactly the BH mass in the metric. Then one can show explicitly that integrating LHS of the latter relation, i.e. $R$, over a 3-dimensional invariant volume one gets a constant finite result which exactly reproduces a constant upon integrating the RHS, i.e. delta-function. To complete this consideration one should also prove that an integration of the LHS and RHS with some test-function $\phi(r)$ belonging to a space of square-integrable functions with a compact support results in $\phi(0)$. This is a routine but straightforward exercise.

From here we learn that the SBH solution mathematically yields $R\propto \delta^3(r)$
which immediately prompts for many questions if one tries to substitute it into equation (\ref{EOMtrace}). One has to define properly what it means to sum up infinite series of all and infinite derivatives of a delta-function and also define their products. Colombeau algebras \cite{Colombeau1992} is a tool to multiply distributions but here we encounter infinite sums of products of various distributions --- a very complicated and not well-understood structure from a mathematical point of view.

To see further that infinite derivatives may give surprises when working with point-support distributions let's resort to a known example \cite{10.1215/S0012-7094-62-02929-0}. Consider an expression
$$e^{a\pd_x^2}\delta(x)$$
on a line $x$ for $a>0$. It is the inverse of the Weierstrass transform of a delta-function. By invoking a Fourier transform one comes back to a Weierstrass transform of a unit
$$e^{a\pd_x^2}\delta(x)=e^{a\pd_x^2}\frac 1{\sqrt{2\pi}}\int dk e^{ikx}
=\frac 1{\sqrt{2\pi}}\int dk e^{-ak^2}e^{ikx}=\frac 1{\sqrt{2a}}e^{-x^2/(4a)}$$
This manifestly shows that an infinite-derivative operator smears a point-support function and provides some everywhere non-zero RHS. On the one hand, this is what we very much want from infinite-derivative models since they are promoted to be UV complete. On the other hand, this is an explicit demonstration that the logic is broken here, as long as the SBH solution is under consideration at least. Indeed, we have started with the idea that there is a point-like mass in the RHS of equations, our solution naively corresponds to a point source with otherwise an empty space but we have ended up not only with a non-point-like RHS but also with a non-compact matter distribution.

To overcome the aforementioned difficulties we exercise below a physical approach to this problem. A regularization will be proposed to make the SBH solution smooth and a relevant part of the RHS of equation (\ref{EOMtrace}) will be computed. It will be shown that regularization cannot be removed in a higher (with more that four) or an infinite-derivative case (apart from some peculiar and most likely unphysical situations to be discussed separately).

\section{Exponential regularization of Schwarzschild metric}

First, let us try to understand what and how should be regularized. Metric is a tensor and as such cannot be said whether it is regular or singular as it is subject to coordinate transformations. A BH horizon is a standard example of a coordinate singularity that can be removed by choosing a different coordinate system. Scalar invariants, however, are not changing with coordinate transformations and are good indicators of real singularities. $R$ would be a nice example but it does not work well exactly for the SBH as its value at zero is not well defined. Another widely used option is the Kretschmann invariant
$$K=R_{\alpha\beta\mu\nu}R^{\alpha\beta\mu\nu}.$$
It touches all the elements of the Riemann tensor. To be absolutely rigorous, its regularity does not guarantee that a manifold is regular but its singularity definitely says that the manifold has a real singularity. For the SBH one can readily get $K=48G^2M^2/r^6$ pointing to a real singularity at $r=0$ and showing a totally fine behavior at the horizon, as expected.

One can proceed by regularizing a metric in an attempt to remove singularity, taking a regularization from  \cite{Bonanno:2000ep} as an example. Consider a function $A(r)$ which defines metric (\ref{startmetricagain}) as follows:
\begin{equation}
    A_{cubic}(r)=1-\frac{2GMr^2}{r^3+r_0^3}
    \label{fcubic}
\end{equation}
This expression restore the SBH behavior for large values of $r$ and its $K$-invariant is
$$K=48G^2M^2\frac{r^{12}-4r^9r_0^3+18r^6r_0^6-2r^3r_0^9+2r_0^{12}}{(r+r_0)^6(r^2-rr_0+r_0^2)^6}$$
which is an expression clearly regular everywhere. However, one can compute using $A_{cubic}(r)$:
$$
K_2=R_{\alpha\beta\mu\nu}\Box^2R^{\alpha\beta\mu\nu}=\frac{576G^2M^2}{r(r+r_0)^{12}(r^2-rr_0+r_0^2)^{12}}\times P_2(r)
$$
where $P_2$ is regular everywhere and $P_2(0)\neq 0$. We see that this expression which is still a scalar invariant is singular.\footnote{$
K_1=R_{\alpha\beta\mu\nu}\Box R^{\alpha\beta\mu\nu}$ turns out to be regular everywhere.
}

This brings forward the following logic. It may happen that we cannot regularize all scalar invariants. Moreover, it is in fact not needed. Regularization (\ref{fcubic}) is great in the framework of GR just because Einstein equations do not involve terms with more than 2 derivatives and testing Kretschmann invariant is simply enough to guarantee that all the terms which one would encounter upon some computations are regular. This is not so in the case of more and needless to say an infinite number of derivatives. To make sure that all the terms in infinite sums of our equation (\ref{EOMtrace}) are regular one has to come up with a regularization which makes function $A(r)$ in the metric exactly unit at the origin and all its derivatives zero. This is necessary to make the RHS, i.e. the matter density distribution regular. It is not guaranteed to be sufficient because a summation of all the terms must be studied as well and amounts to a big deal of our analysis below.

Regularity of every term in equation (\ref{EOMtrace}) can be achieved by means of an exponential regularization and a neat example is as follows:
\begin{equation}
    A(r)=1-\frac{2GM}r e^{-\alpha {r^{-p}}},\quad p>0.
\label{fexp}
\end{equation}
Here $\alpha$ is a regularization parameter. The limit $\alpha\rightarrow 0$ removes regularization and restores the SBH solution. For large $r$ we obviously restore the SBH geometry.

Such an exponential regularization has chances to rectify another weak point of a power-law regularization. Namely, a power-law regularization like (\ref{fcubic}) results in a power-law matter density distribution which seems to be not suppressed enough to be attributed to a ``compact object'' in an ``empty space''. This issue will be addressed in forthcoming publications.

\section{Regularization at work}

\subsection{Schwarzschild Black Hole in GR}

To see how a regularization works let us consider a regularized function $A(r)$ defined such that
\begin{equation}
    A(r)=1-\frac{2GM}r\tilde A(r)
\end{equation}
and also $\tilde A (\infty)=1$ and $\lim\limits_{r\to0}\tilde A(r)/r=0$. The latter limit implies that near $r=0$ we have $\tilde A(r)\approx r^{1+\epsilon}$ with $\epsilon>0$ and thus $\tilde A(0)=0$. These are requirements following from large distance asymptotics and from the fact that $\tilde A(r)$ should regularize our spacetime. Then we can find that a scalar curvature has a neat expression
$$R=\frac{2GM}{r^3}(r^2\tilde A'(r))'.$$
If we work with GR equations than the trace equation gives us
$$M_P^2R=-T$$
In order to test what is the object represented by this stress energy tensor we will integrate both sides over the 3-dimensional space.
\begin{equation}4\pi M_P^2\int_0^\infty Rr^2dr=M\left.\left[r\tilde A'(r)+\tilde A(r)\right]\right|_0^\infty= -E\end{equation}
where $4\pi$ comes from the integral over angular variables and we used that $8\pi G M_P^2=1$. $E$ is the total energy of our matter and the minus sign is coming from the fact that it is an integral of a trace of the energy-momentum tensor, i.e. one index up and one down, while a positively defined energy component is $T_{00}$ with both indexes down. For a static configuration this is essentially a mass of an object. For a compact object it must coincide with the mass parameter $M$ in the metric as otherwise the Newtonian limit is not restored. In what follows we will use the terms total energy and mass interchangeably.

According to our conditions on $\tilde A(r)$ the second term in brackets gives unit. The first term is not such transparent. At $r=0$ it tends to zero because $\tilde A'(r)\approx(1+\epsilon)r^\epsilon$. At $r=\infty$ we demand $\tilde A(r)$ tends to unit and thus its derivative tends to zero faster than $1/r$ (as $1/r$ is a derivative of $\log(r)$ which grows for large $r$). So, we can conclude that in total the first term in brackets results in zero. Hence, we get that LHS corresponds to a constant total energy of matter, or mass, of the SBH and it is exactly the mass parameter $M$ in the metric.

This statement is independent of the form of the regularization function $\tilde A(r)$. This proves that the SBH is a solution to Einstein equations with a physically justified matter distribution.

\subsection{Schwarzschild Black Hole in higher and infinite derivative gravity}

Now we want to do the same for the complete trace equation (\ref{EOMtrace}). Namely we want to compute the total energy of the matter distribution and test its viability. It is possible to massage corresponding expression before picking any metric. On the way of computing the total energy we encounter many terms which look like a total derivative
$$\int d^3x \nabla_\mu j^\mu$$
where we use $\nabla_\mu$ for a covariant derivative. This however does not immediately mean that this expression is zero as we integrate over a 3-dimensional space only. The trick is to write the above expression as an integral over a whole 4-dimensional space-time and divide by the length of the time interval. Then we have
$$\int d^3x \nabla_\mu j^\mu=\lim_{\Delta t\to \infty}\frac 1{2\Delta t}\int_{-\Delta t}^{+\Delta t}dt
\int d^3x \sqrt{-g}\nabla_\mu j^\mu$$
which is a valid transformation due to the fact that the Schwarzschild metric in static coordinates does not depend on time and the integral over $dt$ decouples, and also because in our analysis we have $\sqrt{-g}=r^2$ since a function in front of $dt^2$ and an inverse function in front of $dr^2$ are the same. Being barely a trick it allows now to move derivatives freely and drop total derivative terms.

Considering terms in (\ref{EOMtrace}) we see that, for instance, $-6\lambda\Box\Fc_1(\Box)R$ can be dropped and $\Delta_{2,4}$ can be dropped as well. Moving to a term with ${\Lc_1}^\mu_\mu$ and $\bar \Lc_1$ we see that
\begin{equation}
\begin{split}
&-\lambda\int d^4x\sqrt{-g}({\Lc_1}^\mu_\mu +2\bar\Lc_{1})\\
=&-\lambda\sum_{n=1}^\infty
{{ f}_1}{}_n\sum_{l=0}^{n-1}\int d^4x\sqrt{-g}\left(\pd^\mu  R^{(l)}  \pd_\nu  R^{(n-l-1)}+2R^{(l)}    R^{(n-l)}\right)\\
=&-\lambda\sum_{n=1}^\infty
{{ f}_1}{}_n\sum_{l=0}^{n-1}\int d^4x\sqrt{-g}R^{(l)}    R^{(n-l)}
=-\lambda\sum_{n=1}^\infty
{{ f}_1}{}_n\sum_{l=0}^{n-1}\int d^4x\sqrt{-g}R    R^{(n)}\\
&=-\lambda\int d^4x\sqrt{-g}R\Box\Fc_1'(\Box)R
\end{split}
\end{equation}
Here a prime over a form-factor indicates a derivative with respect to an argument and should not create a confusion with a derivative with respect to $r$ in other instances as the underlying meaning is self-explanatory. Moving from the second to third lines we moved derivatives and removed total derivative terms. The last transformation uses the fact that $\sum\limits_{n>0}n f_{1 n} R \square^n R=R \square \Fc_1'(\square) R$. Continuing the same way for ${\Lc_{2,4}}^\mu_\mu$ and $\bar \Lc_{2,4}$ and canceling redundant time integration with the limits in $\Delta t$ used solely to simplify our computations we arrive at a very compact expression
\begin{equation}
    M-4\pi\lambda\int_0^\infty \left(R\square \Fc'_1(\square)R+L_{\alpha\beta} \square \Fc'_2(\square)L^{\alpha\beta}+W_{\alpha\beta\gamma\delta} \square \Fc'_4(\square)W^{\alpha\beta\gamma\delta}\right)r^2dr=-E.
    \label{energyint}
\end{equation}
We stress that this expression is obtained without specifying a metric at all. The only fact used intrinsically is the asymptotic flatness at large distances which allows to drop boundary terms during integration.

The first term $M$ comes from considering the Einstein-Hilbert term in the action. Other terms represent new physics and we see that a crucial role is played by the derivatives of the form-factors. In particular in the local quadratic gravity where $\Fc_i=\const$ these terms will not contribute to a matter content corresponding to actually any asymptotically flat metric. Hence, the SBH is a physical solution in the fourth-order gravity representing a singular static BH with mass $M$.

In the case of the SBH higher order in derivatives terms may lead in the limit $\alpha\to 0$ to: (i) no contribution and then the corresponding object will have mass $M$ like the SBH would have; (ii) a finite contribution and then the mass of the corresponding object will get corrected with respect to an expected value from the form of the metric -- a very confusing possible outcome as in this case the long distance asymptotics is not restored while everything looking regular otherwise; (iii) a singular contribution which corresponds to an infinite mass of an object and this means that the SBH solution is not compatible with the model.

In our model (\ref{properaction}) we expect form-factors to be analytic at zero and thus $\Fc'_i$ is analytic as well leading to a Taylor expansion
$$\Box\Fc'_i(\Box)=\sum_{n>0}{\hat {f_i}}_n\Box^n,\quad \text{such that}\quad {\hat {f_i}}_n=n{f_i}_n,\quad n>0$$
A term without d'Alembertians under the integral in (\ref{energyint}) can appear only if an original form-factor has a logarithmic contribution. It does not correspond to our initial condition that form-factors are analytic at zero. This is however a very probable additional term upon accounting one-loop counter-terms in the full propagator. To account such a contribution we have to introduce non-zero coefficients $\hat {f_i}_0$. Also for brevity we will rewrite (\ref{energyint}) as follows
\begin{equation}
    M-4\pi\lambda\int_0^\infty \left(R\hat\Fc_1(\square)R+L_{\alpha\beta}  \hat \Fc_2(\square)L^{\alpha\beta}+W_{\alpha\beta\gamma\delta} \hat \Fc_4(\square)W^{\alpha\beta\gamma\delta}\right)r^2dr=-E.
    \label{energyinthat}
\end{equation}
where we introduce new notations for operator functions assuming they contain a constant term in their Taylor expansion at zero:
\begin{equation*}
    \hat\Fc_i=\sum_{n\geq0}{\hat {f_i}}_n\Box^n=\hat {f_i}_0+\sum_{n>0}{n {f_i}}_n\Box^n
\end{equation*}
We may also occasionally name them form-factors specifying explicitly their relation to the expression for the total energy.

Arranging terms with respect to the power of the d'Alembertian we can write the result of integration as
\begin{equation}
    M-4\pi\lambda(E_0+E_1+E_2+\dots)= -E
    \label{infEseries}
\end{equation}
where terms $E_n$ correspond to the terms containing $\Box^n$ in the integral for the total energy (\ref{energyinthat}) and dots represent an integration result for higher order in derivatives terms.

To move further we must use an explicit metric. Computations for a general regularization $\tilde A(r,\alpha)$ seem unfeasible and we proceed by using regularization (\ref{fexp}). In the case $p=1$, the following contributions to the total energy can be computed (hereafter we put $G=1$):
\begin{eqnarray}
E_0&=&\frac{3M^2}{8\alpha^3}\left(60 {\hat{f_1}}_0+7 {\hat{f_2}_0}+4 {\hat{f_4}_0}\right)\\
    E_1&=&\frac{5 M^2}{157464 \alpha ^6} (420 {\hat {f_1}_1} (7168 M-59049 \alpha )
    +{\hat {f_2}}_1 (494592 M-3365793 \alpha )\nonumber\\
    &&\qquad\qquad+44 {\hat {f_4}_1} (7936 M-59049 \alpha ))\\
E_2&=&\frac{2835 \left(8910 {\hat {f_1}_2}+2265 {\hat {f_2}_2}+2213 {\hat {f_4}_2}\right) M^4}{8192 \alpha ^9}\nonumber\\
&-&\frac{143360 \left(25080 {\hat {f_1}_2}+6399 {\hat {f_2}_2}+6386 {\hat {f_4}_2}\right) M^3}{177147 \alpha
   ^8}\nonumber\\
   &+&\frac{42525 \left(84 {\hat {f_1}_2}+13{\hat {f_2}_2}+12 {\hat {f_4}_2}\right) M^2}{16 \alpha ^7}
\end{eqnarray}
We see that these contributions are singular in $\alpha\to0$ limit. We also note that a contribution linear in $M$ never appears in $E_n$ because we are evaluating total energy integrating over an infinite space volume.

Obviously, if we have higher than four but only a finite order of derivatives in the original action (\ref{properaction}) this corresponds to a finite number of terms $E_n$, and than the limit $\alpha\to0$ is always singular leading to an infinite total energy, i.e. an infinite mass. On this basis, we conclude that the SBH is not a solution in this case. The same is true for possible logarithmic terms in form-factors of our model when contribution $E_0$ appears.

In the next Section, we will study a much more subtle question of whether an infinite sum of $E_n$ in (\ref{infEseries}) can have a regular limit $\alpha\to 0$.

\subsection{Gauss-Bonnet combination and related cancellations}
\label{SecGB}

Here we want to emphasize interesting and physically important cancellations that may take place. 
Consider a Gauss-Bonnet term given by
$$\Gc=R^2-4R_{\mu\nu}^2+R_{\mu\nu\alpha\beta}^2=\frac16R^2-2L_{\mu\nu}^2+W_{\mu\nu\alpha\beta}^2.$$
It is a topological term in four dimensions. Its higher derivative generalization which can be written as
\begin{equation}
\Gc_\Fc=\frac16R\Fc(\Box)R-2L_{\mu\nu}\Fc(\Box)L^{\mu\nu}+W_{\mu\nu\alpha\beta}\Fc(\Box)W^{\mu\nu\alpha\beta}
\label{GB}
\end{equation}
is obviously not a topological term even though all the form-factors used are identical. Such a combination is known to give no contribution to the propagator around Minkowski space-time \cite{Koshelev:2022wqj}.\footnote{Frankly speaking, the reason for this is unclear and deserves a deeper understanding.} What is interesting is that yet another cancellation happens, at least in the case of metric (\ref{startmetricagain}). Namely, one can check that in all the expressions for $E_{0,1,2}$ above terms proportional to $M^2$ cancel out if one imposes ${\hat {f_1}_n}=\frac16\hat {f_n}$, ${\hat {f_2}_n}=-2\hat {f_n}$, ${\hat {f_4}_n}=\hat {f_n}$. This can be explained by the fact that terms proportional to $M^2$ correspond to a propagator in a flat space-time which does not get a contribution in this case. We furthermore obtained that a Gauss-Bonnet-type cancellation happens for an arbitrary value of $p$ in our regularization. However, following the just mentioned argument, it should be a general property for asymptotically flat metrics. This is not enough to make all singular in $\alpha\to 0$ limit terms vanish and higher orders in $M$ will survive.



\section{Total energy $E$ of regularized Schwarzschild Black Hole and its singular limit}

\subsection{Entire functions for quantum gravity} 

A plausible form of the propagator in the model (\ref{properaction}) is given by (\ref{prop}) and we repeat it here for a smoother presentation
\begin{equation}
    \Pi(k^2)=\frac{e^{2\sigma(k^2)}}{k^2},
\end{equation}
Here the crucial ingredient is an exponent of an entire function $\sigma(k^2)$.

Power-counting arguments for the renormalizability of an infinite derivative gravity work only for power-law (and not exponential) asymptotics of the form-factors (at least for Minkowski and Euclidean momenta) \cite{Modesto:2014lga}. As polynomials always have zeros, they produce ghosts and should not be used as form-factors. The only option is to have a form-factor which is an exponent of an entire function. Indeed, entire functions are functions analytic everywhere on the complex plane. Polynomials are the simplest examples of entire functions. As such, exponents of entire functions are specific entire functions that have no zeros on the complex plane (and thus their inverse have no poles and no new particles can be identified).
However, to continue using power-counting arguments one wants to arrange that ${\sigma(k^2)}$ behaves as a logarithm along certain directions in the complex plane. The simplest example is an incomplete Gamma-function of a polynomial with a subtracted logarithmic singularity \cite{Tomboulis:1997gg}:
\begin{equation}
\label{sigmaT}
    \sigma(\square)=\frac{1}{4}\left(\Gamma(0,\square^2)+\gamma_E+\log{\square^2}\right).
\end{equation}
A polynomial inside the incomplete Gamma-function can be of any finite degree but we choose it such that the graviton propagator scales as $1/k^4$ for real $k$ resembling the behavior of the local quadratic gravity \cite{Stelle:1977ry,Stelle:1976gc} but now without a ghost. Also, we add the Euler constant $\gamma_E$ for normalization $\sigma(0)=0$ to restore an IR propagator.

Entire functions are classified by their maximal growth rate which is usually characterized by an order $\rho$ and a type $s$ \cite{levin1996lectures}. An entire function $\phi(z)$ can be represented by its Taylor series at zero
\begin{equation}
    \phi(z)=\sum_{n=0}^{\infty}a_n z^n.
\end{equation}
Recall that, according to the Cauchy–Hadamard theorem, the power series around zero converges inside the circle $|z|<z_r$ with the radius
\begin{equation}
    z_r= \lim_{n\rightarrow\infty}|a_n|^{-\frac{1}{n}}.
\end{equation}
For entire functions, the radius of convergence of its series at zero is $z_r=\infty$. 

An order of the entire function can be defined as 
\begin{equation}
    \rho={\rm max}\left(\lim_{|z|\rightarrow \infty}\frac{\log{(\log{|\phi(z)|})}}{\log{|z|}}\right)=\lim_{n\rightarrow \infty}\frac{n \log{n}}{-\log{a_n}}.
\end{equation}
If $\rho$ is finite and non-zero, a type $s$ of the function is:
\begin{equation}
       s={\rm max}\left(\lim_{|z|\rightarrow \infty}\frac{\log{|\phi(z)|}}{|z|^{\rho}}\right)=\lim_{|z|\rightarrow \infty}\left(\frac{n}{e\,\rho}\right)\,|a_n|^{\frac{\rho}{n}}
\end{equation}
The maximal growth rate of an entire function with order $\rho$ and type $s$ is given by $e^{s \,z^{\rho}}$ for $|z|\rightarrow \infty$.

\begin{figure}
    \centering
    \includegraphics[width=10cm]{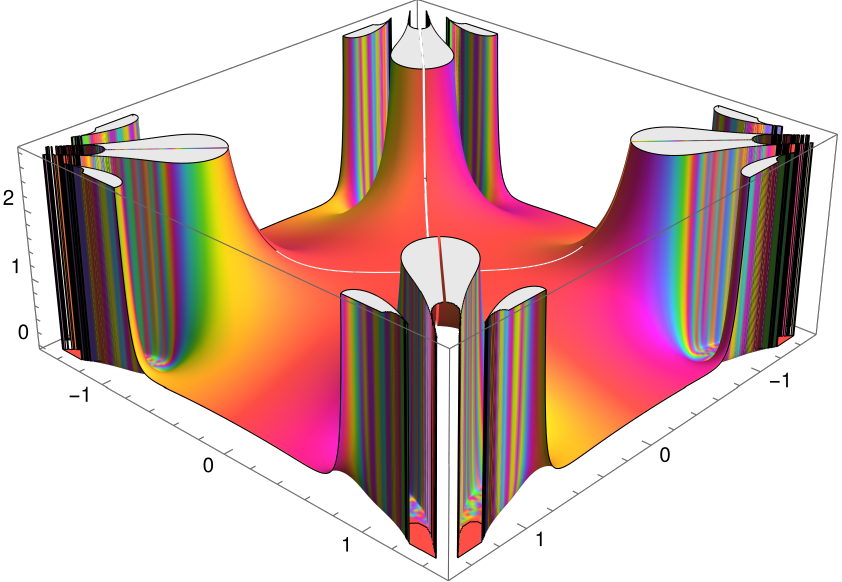}
    \caption{Plot of an absolute value and a phase of the function \eqref{sigmaT}. Real and imaginary axes correspond to the same behavior of graviton propagator decaying as $1/k^4$ for both Minkowski and Euclidean momenta. This is possible only for entire functions of an infinite order with an essential singularity at infinity of the form $e^{e^z}$ which shows up in a messy behavior along some unphysical directions in the complex plane.}
    \label{fig:Tomboulis}
\end{figure}

Polynomials are entire functions of order zero. However, not only polynomials are entire functions of order zero. There are also entire functions that grow faster than polynomials and slower than any exponent, for example, $\phi(z)=\sum\limits_{n\geq0} e^{-n^2} z^n$. Infinite order functions grow as $e^{e^z}$ and faster. Remarkably, function $e^{2\sigma(z)}$ with $\sigma(z)$ given by \eqref{sigmaT} belongs to the infinite order class. We plot this function in Figure~\ref{fig:Tomboulis}.

It is not clear whether the ultimate non-perturbative propagator of a graviton has the same form as the bare propagator for all values of momenta, including its extension to the complex plane \cite{Reuter:2019byg}. The non-perturbative propagator is certainly not an entire function on the complex plane because it must have a branch cut at the one-loop level order. However, we expect that the corrections along the real axis are small, compared to the tree level. Nevertheless, the growth rate of the resummed propagator can well be different. For this reason, it is not clear whether renormalizable gravity still predicts the full graviton propagator to be a function of an infinite order. In the next Subsection, we discuss different possibilities for the order of the form factors in the gravity Lagrangian (\ref{properaction}) and their large-momentum behavior.

\subsection{General properties of series for the total energy $E$}

We proceed by applying the above considerations of entire functions to the series (\ref{infEseries}) representing the total energy of the matter corresponding to our space-time metric. It is the total mass of the object but it is not guaranteed to be value $M$ in the case of the (regularized) SBH metric. Also, it is not guaranteed to be compact unless the sum of $E_n$ vanishes in the limit $\alpha\to 0$. We defer this important question of the mass distribution with respect to the radius to future studies.

Hereafter we use the regularization of the metric (\ref{fexp}) with an arbitrary positive parameter $p$.
Also, we put $G=1$ for simplicity.
We start by noting, that having terms like $1/\alpha^k$ arranged in an infinite series does not immediately mean that the limit $\alpha\rightarrow 0$ is divergent upon summation. For example,
\begin{equation}
\label{resum_exp}
    \sum_{k=0}^{\infty}\frac{(-1)^k}{k!\,\alpha^k}=e^{-1/\alpha}\xrightarrow{\alpha\to0} 0, \quad \alpha>0.
\end{equation}
In this case, the series can be summed up and lead to zero if the limit is performed after the summation.
Series with a similar behavior would give no contribution to the total energy.
However, this is possible only if the series has an infinite radius of convergence.

If the radius of convergence is finite, it means that the theory allows for regular solutions for the metric corresponding to a compact regular object of a finite mass looking similarly to a BH from a large distance. However, it is not possible to take a limit $\alpha\to 0$ keeping the total energy of the matter distribution (i.e. mass of this object) finite. Thus, no singular solutions are possible in this case, as well as in the case when the convergence radius is zero. 

In the case when the radius of convergence of the series is infinite and the summation result is zero in the limit $\alpha \to 0$ its contribution to the total energy vanishes. In this case, the singular Schwarzschild solution smoothly goes at least through the trace of the equations of motion with infinite derivatives. This is when the familiar singular BH solution is also an exact solution to the equations of motion in a non-local gravity. However, we will show in the subsequent analysis that this never happens in a perturbative and ghost-free theory which requires at least exponential form-factors or better their structure is like suggested in \cite{Tomboulis:1997gg}, see Eq. \eqref{sigmaT}.

An intermediate case when the series is summed up to a finite constant would correspond to a strange configuration when a large distance asymptotic is not recovered as the total mass of the object does not correspond to the mass parameter $M$ in the metric, while the total energy is finite. Also, it is not obvious how to construct an example of such a series.

\subsection{Convergence analysis}

\begin{figure}[h!]
        \centering
        \includegraphics[width=2.9in]{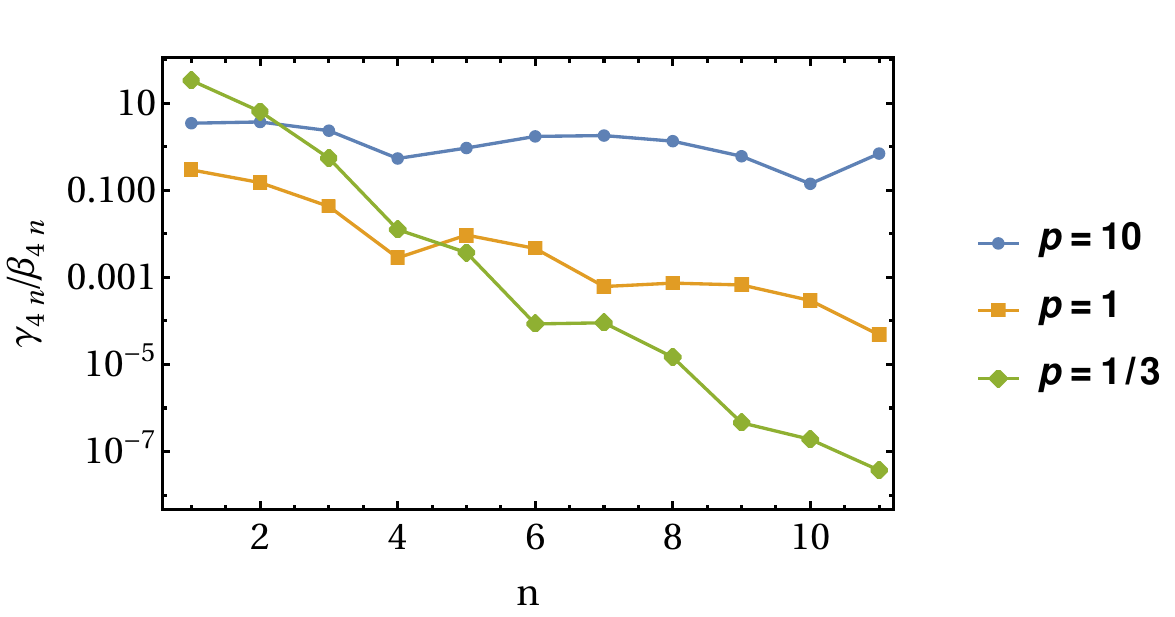}
        \includegraphics[width=2.9in]{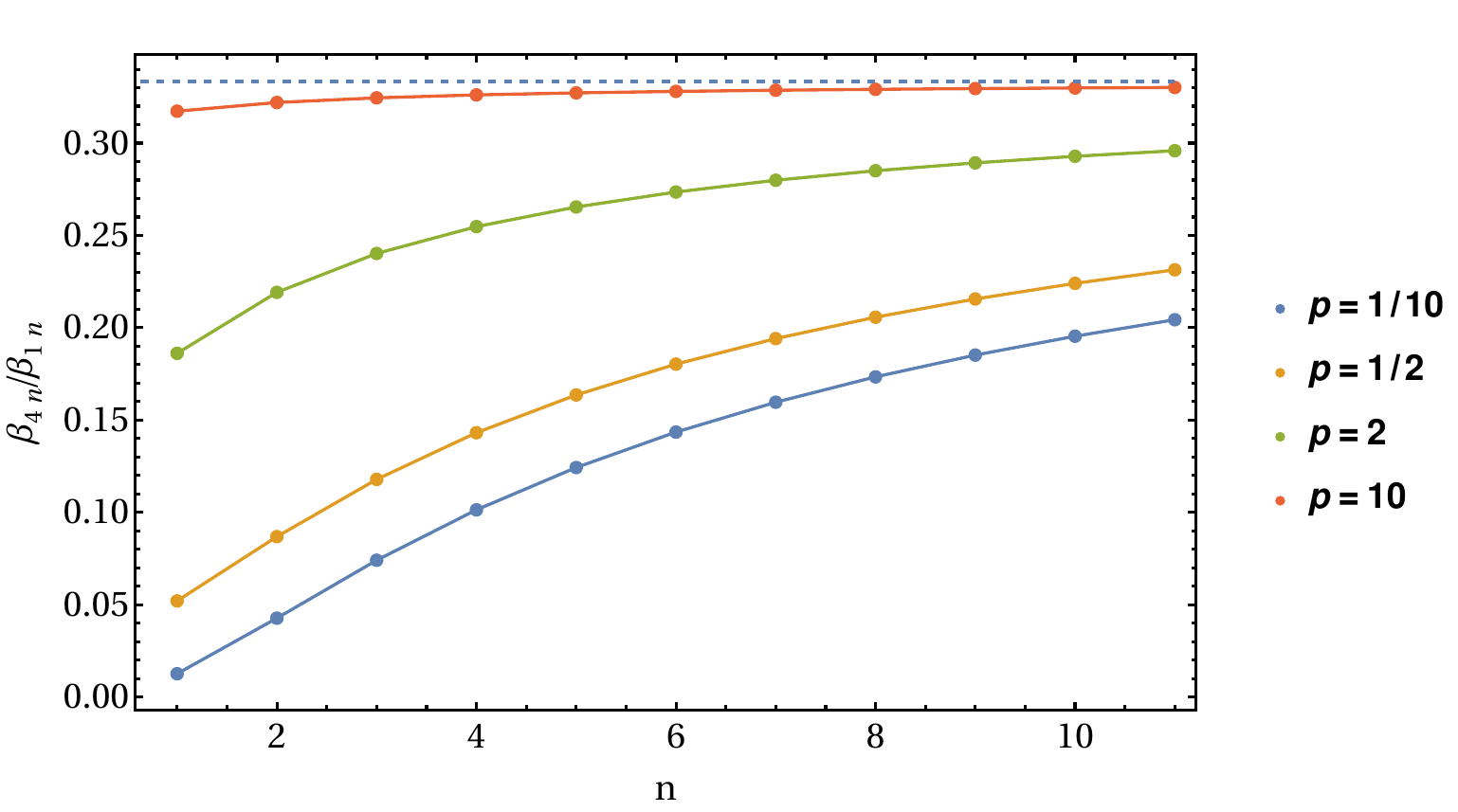}
        \caption{The left plot shows the fraction ${\gamma_4}_{n}/{\beta_4}_{n}$ for different values of $p$. It illustrates that the series coming in front of $M^3$ have either the same or even better convergence properties compared to the series coming in front of $M^2$ because this ratio is always decreasing with $n$. This is why it is enough to study the large $n$ asymptotics of the series in front of $M^2$. The right plot shows the fraction ${\beta_4}_{n}/{\beta_1}_{n}$ for different values of $p$. We can see that for large $p$ the ratio clearly approaches $1/3$. One might expect the cancellations in case if ${\hat{f_4}}_n=-3 {\hat{f_1}}_n$ but we checked that such cancellations do not change the leading factorial asymptotics (\ref{largep}).}
        \label{plotg}
    \end{figure}

We perform a brute force computation of the total energy of the matter distribution corresponding to metric (\ref{startmetricagain}) with a regularized function $A(r)$ given by (\ref{fexp}) up to the terms containing 11-th power of the d'Alembertian.
It is instructive to expand the resulting expression in powers of the mass parameter $M$.
Schematically the resulting expression looks as follows
\begin{equation}
\label{trace_final}
\begin{split}
    &M-4\pi\lambda M^2 (2\alpha)^{-\frac{3}{p}}\left[\left(\sum_{n=0}^{\infty}(-1)^n{\beta_{1}}_n(p)\hat{f_{1}}_n(2\alpha)^{-\frac{2 n}{p}}\right)+\left(\sum_{n=0}^{\infty} (-1)^n{\beta_{2}}_n(p)\hat{f_{2}}_n(2\alpha)^{-\frac{2 n}{p}}\right)+\right.\\
  &  \left.\qquad\qquad\qquad\quad+\left(\sum_{n=0}^{\infty}(-1)^n{\beta_{4}}_n(p)\hat{f_{4}}_n(2\alpha)^{-\frac{2 n}{p}}\right)\right]-\\
  &-4\pi\lambda M^3 (2\alpha)^{-\frac{6}{p}}\left[\left(\sum_{n=1}^{\infty}(-1)^{n-1}{\gamma_{1}}_n(p)\hat {f_{1}}_{n}(2\alpha)^{-\frac{2 (n-1)}{p}}\right)+\left(\sum_{n=1}^{\infty} (-1)^{n-1}{\gamma_{2}}_n(p)\hat{f_{2}}_{n}(2\alpha)^{-\frac{2 (n-1)}{p}}\right)+\right.\\
  &  \left.\qquad\qquad\qquad\quad+\left(\sum_{n=1}^{\infty}(-1)^{n-1}{\gamma_{4}}_n(p)\hat {f_{4}}_{n}(2\alpha)^{-\frac{2 (n-1)}{p}}\right)\right]+O(M^4)=-E
    \end{split}
\end{equation}
Here we have defined ${\beta_i}_n$ to be the coefficients for the $M^2$ part of the expansion
and ${\gamma_i}_n$ for the $M^3$ part of expansion respectively. These are essentially combinations of powers, factorials, and polynomials in $p$. Full expressions up to $n=5$ are listed in Appendix~\ref{appbetagamma}.
{We study the properties of the obtained coefficients by fitting the leading behavior of each term with some function of $p$ and $n$.} It turns out that the series in front of $M^3$ and higher powers of $M$ are subdominant, compared to series in front of $M^2$ (see Figure \ref{plotg}, left plot), in the sense that the series in front of powers of $M$ greater than 2 will converge for granted if the series in front of $M^2$ is convergent. For this reason, we mainly study the contribution proportional to $M^2$.

\begin{figure}[h!]
        \centering
        \includegraphics[width=2.9in]{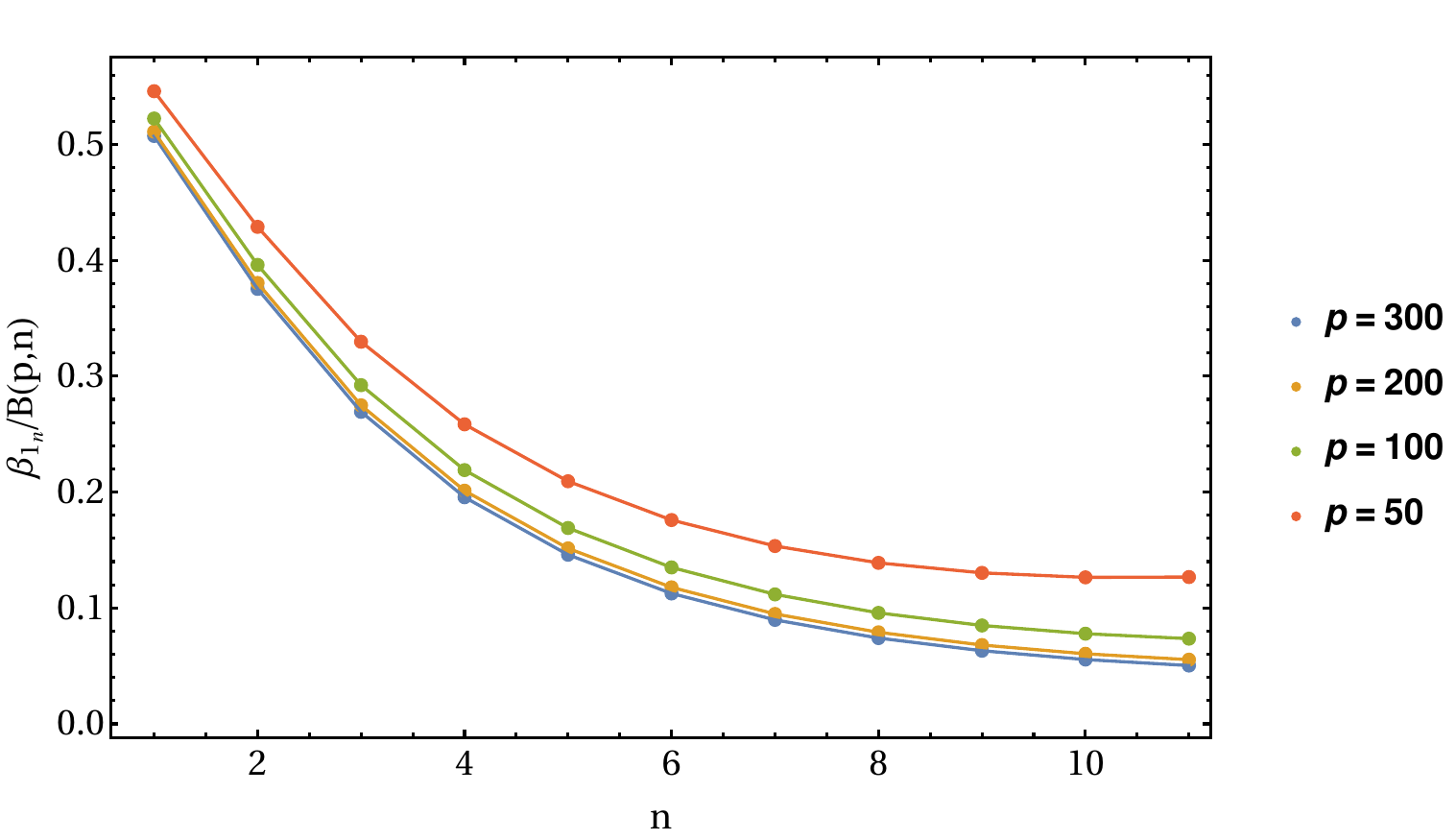}
     \includegraphics[width=2.9in]{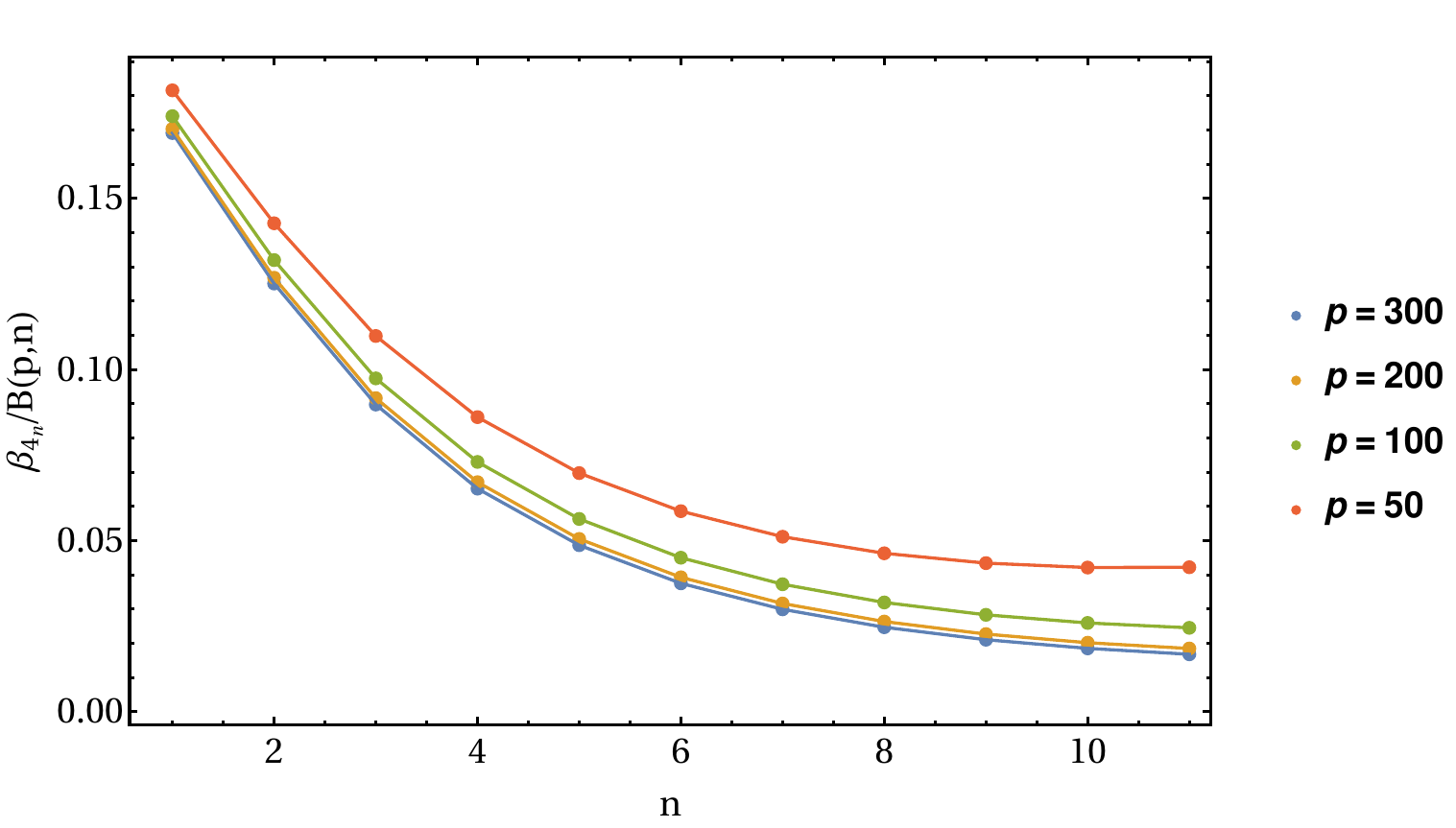}
      \includegraphics[width=2.9in]{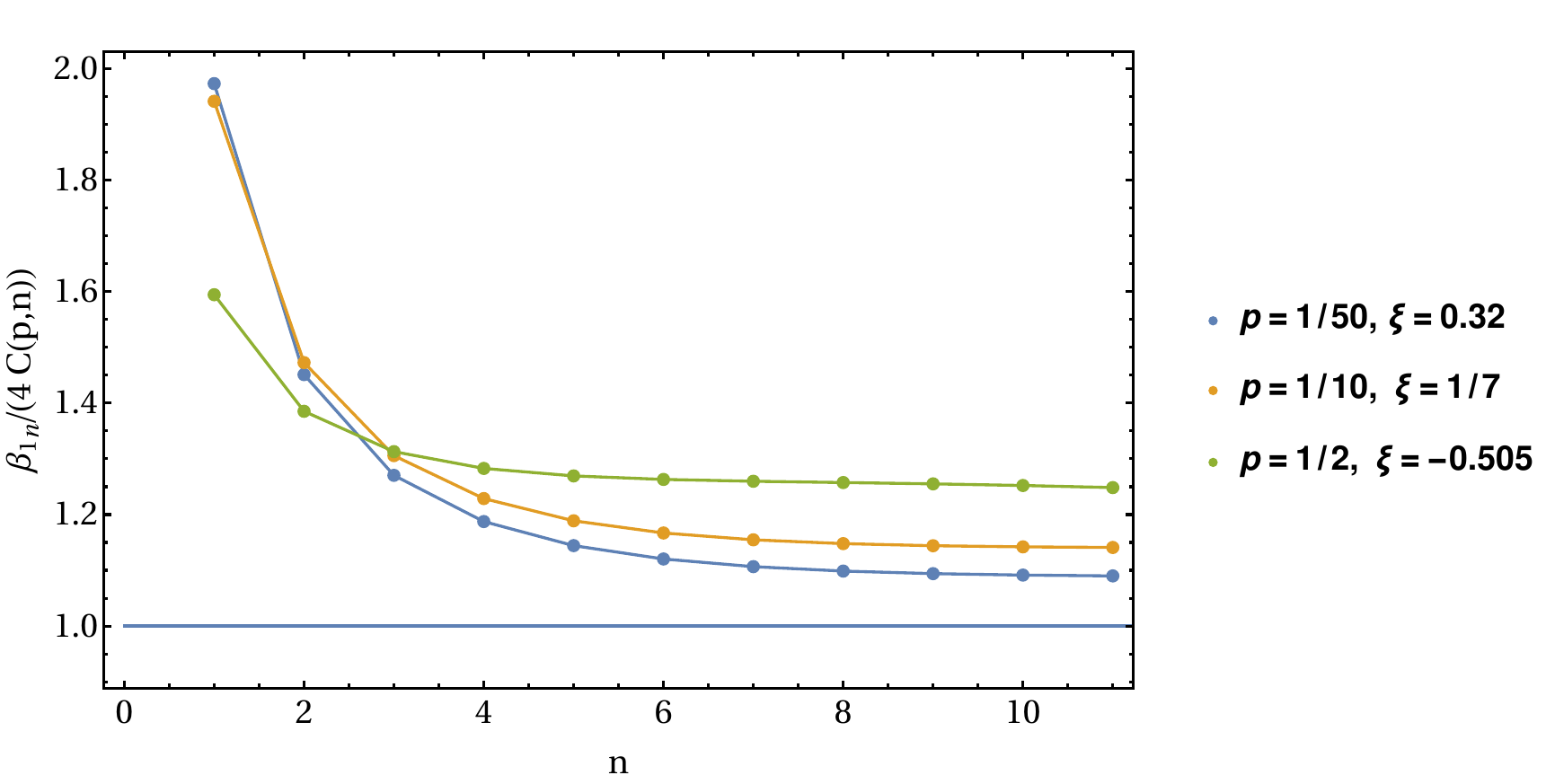}
       \includegraphics[width=3in]{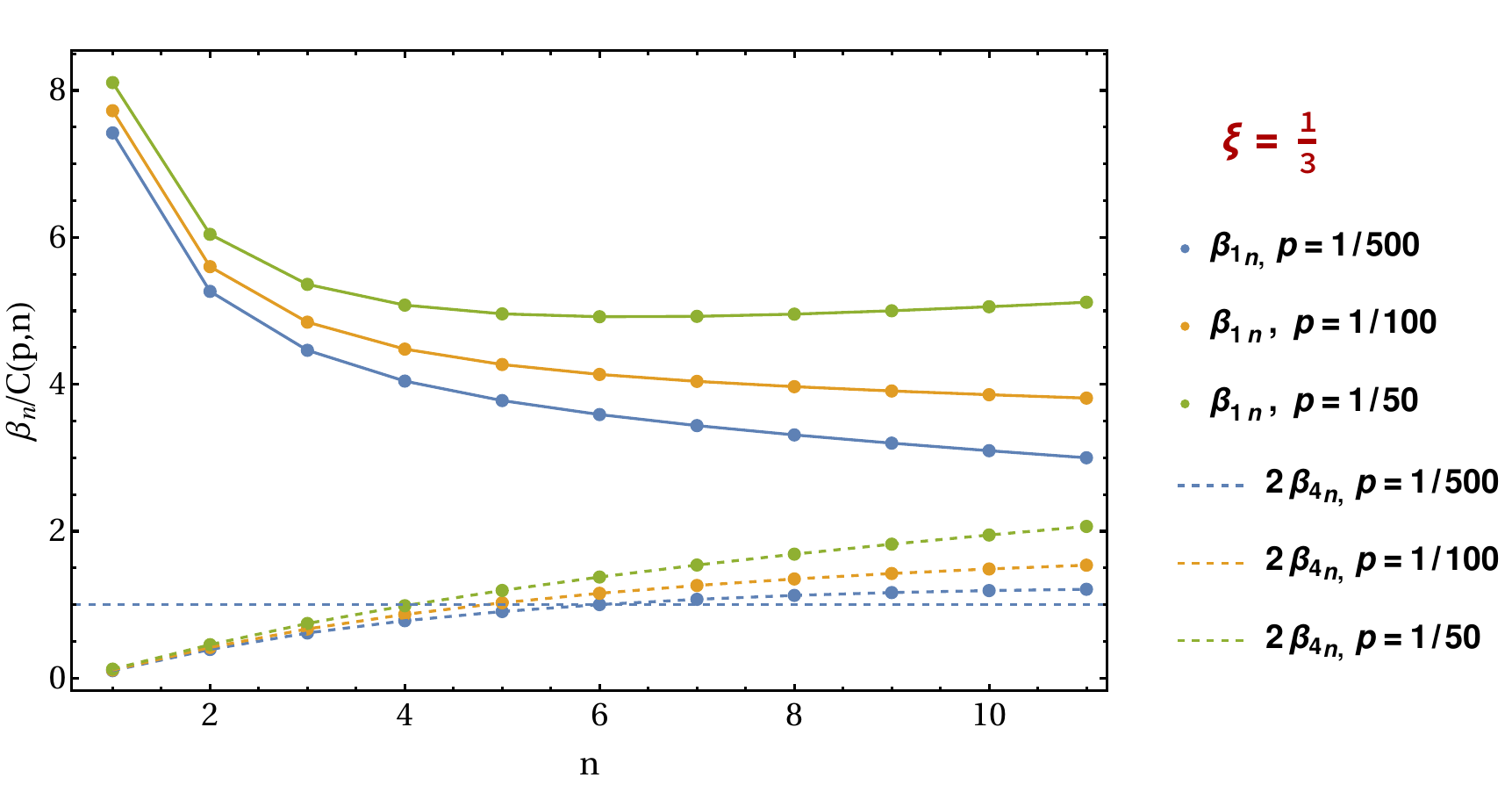}
        \caption{The plots show the first 11 series coefficients $\beta_n(p)/B(n,p)$, $\beta_n(p)/C(n,p)$ where $B(n,p)$ and $C(n,p)$ are written in Eqs. \eqref{largep} (large $p$), \eqref{smallp} (small $p$). The upper left plot illustrates Eq. \eqref{largep} for ${\beta_1}_{n}$, and the upper right plot illustrates ${\beta_4}_{n}$. The lower plots illustrate the limit of small (left) and very small (right) $p$ for ${\beta_1}_{n}$. We checked that ${\beta_4}_{n}$ has the same behavior. We can see that fitting $\xi=1/3$ in Eq. \eqref{smallp} works well only for tiny $p$, as the growth of ${\beta_1}_{n}$ is not compensated even for $p=1/50$ (green solid line). However, a lower value of $\xi$ provides a better fit for non-vanishing $p$ (lower left plot). } 
        \label{plots}
    \end{figure}

Analyzing the behavior of the terms ${\beta_1}_n(p),~{\beta_2}_{n}(p),~{\beta_4}_{n}(p)$ we find the following leading order approximations for the coefficients as functions of $p$ and $n$ at large $p$,
\begin{equation}
\label{largep}
    {\beta_{i}}_n(p)\propto O(1)B(p,n),\quad B(p,n)= p^{2n+1}\,\Gamma\left(2+\frac{2n+1}{p}\right)\Gamma\left(\frac{3 n}{2}\right).
\end{equation}
where $i=1,2,4$ and the $O(1)$ factor is different for different $i$. The large $p$ asymptotics are depicted in Figure \ref{plots} (upper left and upper right plots)

For small $p$ we obtained a different asymptotic behavior,
\begin{equation}
\label{smallp}
    {\beta_{i}}_n(p)\propto O(1)C(p,n),\quad C(p,n)= p\,\Gamma\left(2+\frac{2n+1}{p}\right)\Gamma(2 n)\,e^{-(1+\xi(p))},
\end{equation}
where $i=1,2,4$, the $O(1)$ factor is different for different $i$, and $\xi(p)$ varies from $-1/2$ for $p=1/2$ to $1/3$ for $p\rightarrow 0$. These asymptotics are illustrated in Figure \ref{plots} (lower left and lower right plots)

This yields a relation between the total energy of the object described by our solution and coefficients $\hat {f_i}_{n}$ in the Taylor expansion of the operators in (\ref{energyinthat}). Notice that ${\beta_i}_{n}$ typically grows very fast with $n$ which means that resummation of the series to a finite total mass is possible only if $\hat {f_i}_n$ are decreasing very fast. To be precise, the convergence of the series for $\alpha\ne 0$ and any value of $p$ is granted only if
\begin{equation}
\label{fnLimit}
    \lim_{n\rightarrow\infty} \frac{|\hat {f_i}_{n}|}{e^{q n \log{n}}}=0
\end{equation}
for any value of $q>0$. This is possible, for example, if $\hat {f_i}_n\propto e^{-n^2}$, but it leads to strong constraints on the behavior of form-factors at large momenta. Namely, the rate of decreasing of the coefficients in series is directly related to the order parameter of the entire function, due to Cauchy-Hadamard theorem. The behavior \eqref{fnLimit} means that form-factors leading to convergent series independently of the regularization can only be the order-0 entire functions.

For exponential-type form-factors one can still adjust the large enough value of $p$ leading to convergent series for a non-zero range of $\alpha$. Eq.\eqref{fnLimit} tells us that it is possible if form-factors have order parameter $\rho\leq3/2$ (i.e. they grow at most as $e^{\square^{3/2}}$ in some directions on the complex plane). However, this possibility is regularization-dependent which means that $E$ doesn't have a well-defined unique limit for $\alpha\rightarrow 0$. Most likely, this situation should be interpreted as the possibility of having a solution with singularity for collapsing matter if one approaches it in a very special way. This is still very far from a guaranteed presence of a singularity in physically justified BH-s. 

Finally, let us recall here that the behavior of form-factors expected from a theory that has chances to be perturbatively renormalizable and unitary is limited to operator functions with power-law asymptotics for certain directions in the complex plane, like (but not limited to) function (\ref{sigmaT}). This is possible only if the form-factors have order parameter $\rho=\infty$, which means the maximal growth as $e^{e^{\square}}$. By definition of the order, their Taylor coefficients decay slower than $e^{q n\log{n}}$ with an arbitrary $q$, including $q=3/2$ which is the boundary of allowed behavior needed for convergence (see \eqref{fnLimit}).
Hence, functions of an infinite order fall into a class of functions for which resummation of series \eqref{trace_final} cannot be done. They are divergent for any $\alpha\ne 0$, and this property does not depend on the value of parameter $p$. This is what brings us to a far-going conclusion that the form-factors expected in quantum gravity do not allow singular spherically symmetric solutions with finite total energy.
{In principle, one can imagine some tuning of coefficients ${\hat{f_i}}_n$ which improves the convergence of the series. For instance, we have found that for large $p$ coefficients ${\beta_i}_n$ tend to be related as ${\beta_1}_n\rightarrow 3 {\beta_4}_n$ as $n$ grows, see Figure\ref{plotg}, right plot. It is interesting to mention that this ratio can lead to cancellations if ${\hat{f_4}}_n=-3 {\hat{f_1}}_n$ which exactly corresponds to the unitarity condition \eqref{Ucond} when the theory has the only massless spin-2 degree of freedom. However, we found that no drastic cancellation happens, and the leading factorial behavior \eqref{largep} is not weakened.}

Even though the above considerations were done for the series in the integral (\ref{energyinthat}) for the total energy $E$ all the results obtained for form-factors $\hat\Fc_i$ can be readily transferred to form-factors $\Fc_i$ in the model Lagrangian (\ref{properaction}). This is because form-factors with and without a hat are connected by a differentiation and by a multiplication by an argument --- operations that do not change neither a growth rate nor a type of an entire function.

\section{Conclusions and discussion}

In this work, we have studied whether the presence of higher and infinite derivative terms in the gravity Lagrangian is compatible with singular BH solutions. In our analysis, we used a regularized metric which approaches the Schwarzschild metric when regularization is removed. Since we need to make regular all terms in the equations of motion, including infinite derivatives of the Riemann tensor, a power-law regularization is not satisfactory. We are using an exponent of an inverse power of radius $r$ instead and this perfectly does the job of making all terms in the EOM-s with an arbitrary number of derivatives regular. This regularization also allows us to compute analytically the contribution of each term in the trace of the generalized Einstein equations to the total energy by integrating over the whole 3-dimensional space. In a static configuration, the total energy is just a mass of a centrally symmetric matter distribution. The result is a series of terms corresponding to different powers of d'Alembertian operator under the integral. An ability to get analytic expressions at this stage makes it feasible to conclude about the convergence properties of this series. 

We found that this 3-dimensional integral results in an infinite sum containing the Taylor coefficients of the expansion of the model form-factors and inverse powers of the regularization parameter $\alpha$. This series occurs to be divergent for the majority of exponential form factors, thus making it not possible to admit a singular limit $\alpha\to0$ --- the Schwarzschild metric --- to be a solution in an infinite derivative gravity with a finite total mass of the corresponding matter. Our results can be summarized as follows with respect to the growth rate of form-factors: 
\begin{itemize}
    \item Case 1: $|\Fc_{1,2,4}(\square)|$ growth faster than $e^{q\,\square^{3/2}}$ for some $q>0$ in some direction on the complex plane. The contribution of infinite derivative terms requires the infinite total mass of the matter distribution because the corresponding series diverges for any finite $\alpha$. This is the case for any parameter $p$ in the regularised metric. This contradicts the notion of a compact object with a finite mass.
    \item Case 2:  $|\Fc_{1,2,4}(\square)|$ grow slower than $e^{q\,\square^{3/2}}$ for some $q>0$, and $\Fc_{1,2,4}$ have some finite orders $\rho_i\neq0$. The series are either divergent or convergent depending on the parameter $p$. This means that there is a smooth path to the singular Schwarzschild solution with a finite total mass but it is not guaranteed that it is stable and that this kind of solution is realized in Nature. The regularization dependence of the finiteness of the total mass forbids us from concluding whether this is a possible scenario without an additional stability analysis in this case.
    \item Case 3: $\Fc_{1,2,4}(\square)$ are order-0 entire functions whose maximal growth rate is faster than a polynomial and slower than an exponent. In this peculiar case, the series for the total mass is convergent everywhere for any parameter $p$, and, thus, they do sum up to zero or to some finite value, especially if the series is alternating (this is the case when form-factors are growing, and the propagator is decaying). Then the Schwarzschild solution should remain a solution in this theory if the series has a vanishing sum. 
\end{itemize}
Here we want to immediately express our negative stance towards the last case. Recall that if the non-perturbative gravity can be formulated in terms of an effective action, it should be capable of resolving the singularity problem of the BH-s. For this reason, we expect that a good candidate for quantum gravity theory should forbid the singularity in the center of the BH while leaving the large $r$ limit of the Schwarzschild solution untouched. Therefore, we expect that the form factors mentioned in Case 3 are disfavoured. Notice that the loop-corrected action for the pure GR \cite{Donoghue:1995cz} should also lead to the sub-exponential form factors in the UV which perfectly fits with the expectation that the singular SBH is present in that case.

Our results were obtained for a particular choice of the form of regularization. Will they change if we consider another type of regulating function that interpolates between some regular metric and the SBH solution? In fact, we have found that the steepest choice of $p\rightarrow \infty$ leads to the better convergence of the series. One might expect that taking an even steeper function (like ${\exp}(e^{-\alpha/r})$) we can still achieve the convergence for the Case~1 form-factors. As the detailed analysis is much more complicated, we leave this question for future study. However, after a brief analysis of our regularisation, we expect that making the regularization function steeper will not help because taking large $p$ doesn't make the coefficients of the series grow slower than $\Gamma(3n/2)$.

Although we have studied only the trace of the equations of motion and concentrated only on the contribution proportional to $M^2$, our result is quite generic. Namely, we have found that series in front of $M^3,~M^4,~\dots$ converge if the one in front of $M^2$ does.

Also, since we were studying only the total matter-energy integrating over the whole space, we did not receive terms proportional to $M$ and also noticed that the terms proportional to $M^2$ receive the contributions only from the terms in the gravity action which add to the flat space graviton propagator. It was shown \cite{Biswas:2016egy} that all covariant theories of gravity around a flat space-time reduce to the Lagrangian \eqref{properaction} if the consideration is limited to the propagator only. Thus, there is a one-to-one correspondence between the $M^2$ contribution to the total energy of a centrally symmetric matter distribution and the full non-perturbative graviton propagator. Hence, there is a one-to-one correspondence between the graviton propagator and the fact that singular BH-s exist in a theory. 
As a consequence, omitting higher point interactions seems to be a good approximation, at least for deriving our conclusions about the absence (or presence) of singular solutions with a finite total mass.

There are at least two important questions not covered in this paper but which should definitely be studied. The first is the inclusion of other terms in the model action apart from only quadratic in curvature with d'Alemebrtian operators. These terms are not supposed to change the present results as their contributions are expected to be subdominant to the terms studied in this manuscript. This is because these other terms even though will have the same mass dimension as considered here, do have less derivatives acting on a metric at once. A more complete study that explicitly includes all terms of six- and eight-order gravity (like cubic and quartic in curvature and not only those with d'Alembertian operators) in an arbitrary dimension is a matter of a forthcoming paper \cite{AKprogress}.
The second question is the computation of the radial mass distribution of the obtained regular objects in the most plausible situations where there is no singular limit. This may have long propagating consequences including potential measurable effects of new BH mass profiles.
There are surely many other questions including the study of the interior of new BH solutions, for instance. It is also very important to examine the other singular BH candidate solutions, primarily Kerr BH \cite{Kerr:1963ud}.

\acknowledgments
AK acknowledges the great hospitality of Perimeter Institute where during a visit important discussions with Alessia Platania regarding this project took place, long and illuminating discussions with Ivan Kolar regarding Colombeau algebras and related mathematical aspects, and with Oleg Melichev. The authors are thankful to Kellogg Stelle and Tobi Wiseman for fruitful discussions on the subject of this paper.
AT is supported in part by the National Natural Science Foundation of China (NSFC) under Grant No. 12147103. The work of AT at the beginning was supported by STFC grant ST/T000791/1.

\appendix
\section{Equations of motion}
\label{appeom}

Full equations of motion for action (\ref{properaction})
\begin{subequations}
\begin{eqnarray*}
  &-&M_P^2G^\mu_\nu-\Lambda \delta^\mu_{\nu}\label{EHL}\\
&+&\lambda\left[\frac12\delta^\mu_\nu R  \Fc_1(\Box)
R\right.\label{1hR}\\
&+&2(-R^\mu_\nu+\nabla^\mu\pd_\nu-\delta^\mu_{\nu}
\Box) \Fc_1(\Box) R\label{1xR}\\
&+&{\Lc_1}^\mu_\nu
-\frac{1}{2}\delta^\mu_{\nu}\left({\Lc_1}^\sigma_\sigma+\bar\Lc_{1}\right)\label{1ss}\\
&+&\frac12\delta^\mu_{\nu} {L}^\alpha_\beta\Fc_2(\Box){L}^\beta_\alpha-2 {L}^\mu_{\beta}\Fc_2(\Box){L}^\beta_\nu\label{2hLiL}\\
&+&2\left(-\frac12 \delta^\mu_\sigma\delta_\nu^\rho\left(\frac12R+\Box\right)+\delta^\mu_\sigma\nabla^\rho\nabla_\nu-\frac12\delta^\mu_\nu\nabla_\sigma\nabla^\rho\right)\Fc_2(\Box){L}^{\sigma}_{\rho}\label{2xL}\\
&+&{\Lc_2}^\mu_\nu
-\frac{1}{2}\delta^\mu_{\nu}\left({\Lc_2}^\sigma_\sigma+\bar\Lc_{2}\right)+2{\Delta_2}^\mu_\nu\label{2ss}\\
&+&\frac12\delta^\mu_\nu W_{\gamma\delta\alpha\beta}
\Fc_4(\Box)W^{\gamma\delta\alpha\beta}-2W^\mu_{\delta\alpha\beta}\Fc_4(\Box)W_\nu^{\phantom{\nu}\delta\alpha\beta}\label{3hWiW}\\
&+&2\left(R_{\alpha\beta}
+2\nabla_\alpha\nabla_\beta\right)
\Fc_4(\Box)W_{\nu}^{\phantom{\nu}\alpha\beta\mu}\label{3xW}\\
&+&\left.{\Lc_4}^\mu_\nu
-\frac{1}{2}\delta^\mu_{\nu}\left({\Lc_4}^\sigma_\sigma+\bar\Lc_{4}
\right)+4{\Delta_4}^\mu_\nu\right]=-T^\mu_\nu\label{3ss}
\end{eqnarray*}
\label{EOM}
\end{subequations}
where various quantities are defined after Eq. (\ref{properaction})

\section{Polynomial coefficients ${\beta_i}_n$ and ${\gamma_i}_n$}
\label{appbetagamma}
Here we present our results on the expansion coefficients ${\beta_i}_n$ and ${\gamma_i}_n$ up to $n=5$.
\begin{itemize}
\item Series for ${\beta_{1}}_n$.
\end{itemize}
\begin{equation*}
{\beta_{1}}_0=\frac{1}{4} p \left(2 p^2+3 p+25\right) \Gamma \left(2+\frac{3}{p}\right)
\end{equation*}
\begin{equation*}
{\beta_{1}}_1=\frac{1}{4} p \left(16 p^4+70 p^3+175 p^2+250 p+49\right) \Gamma \left(2+\frac{5}{p}\right)  
\end{equation*}
\begin{equation*}
{\beta_{1}}_2=\frac{1}{16} p \left(272 p^6+2212 p^5+7336 p^4+14063 p^3+15015 p^2+5733 p+729\right) \Gamma \left(2+\frac{7}{p}\right) 
\end{equation*}
\begin{equation*}
\begin{split}
&{\beta_{1}}_3= \frac{1}{64} p \left(7936 p^8+99216 p^7+498300 p^6+1362204 p^5+2277129 p^4+2272104 p^3+1155170 p^2+\right.\\
   &\left.+284076 p+27225\right) \Gamma \left(2+\frac{9}{p}\right)
   \end{split}
\end{equation*}
\begin{equation*}
\begin{split}
&{\beta_{1}}_4= \frac{1}{256} p \left(353792 p^{10}+6132896 p^9+43636648 p^8+169875772 p^7+407283954 p^6+632696295 p^5+\right.\\
   &\left.+629270917 p^4+376198438 p^3+130175144 p^2+24132119 p+1863225\right) \Gamma \left(2+\frac{11}{p}\right)
   \end{split}
\end{equation*}
\begin{equation*}
\begin{split}
&{\beta_{1}}_5= \frac{1}{1024} p \left(22368256 p^{12}+504188672 p^{11}+4767870432 p^{10}+25130915160 p^9+82814137068 p^8+\right.\\
   &\left.+181096728306 p^7+269854652171 p^6+272733046690 p^5+181856368551
   p^4+78051710022 p^3+\right.\\
   &\left.+20726847297 p^2+3103980750 p+200930625\right) \Gamma \left(2+\frac{13}{p}\right)
   \end{split}
\end{equation*}

\begin{itemize}
\item Series for ${\beta_{4}}_n$.
\end{itemize}
\begin{equation*}
{\beta_{4}}_0=\frac{1}{12} p \left(2 p^2+3 p+1\right) \Gamma \left(2+\frac{3}{p}\right)
\end{equation*}
\begin{equation*}
{\beta_{4}}_1=\frac{1}{12} p \left(16 p^4+70 p^3+79 p^2+10 p+1\right) \Gamma \left(2+\frac{5}{p}\right)
\end{equation*}
\begin{equation*}
{\beta_{4}}_2=\frac{1}{48} p \left(272 p^6+2212 p^5+6184 p^4+7007 p^3+2991 p^2+693 p+81\right) \Gamma \left(2+\frac{7}{p}\right)
\end{equation*}
\begin{equation*}
\begin{split}
&{\beta_{4}}_3= \frac{1}{192} p \left(7936 p^8+99216 p^7+472188 p^6+1089180 p^5+1290249 p^4+798984 p^3+288962 p^2+\right.\\
   &\left.+60300 p+5625\right)\Gamma \left(2+\frac{9}{p}\right)
   \end{split}
\end{equation*}
\begin{equation*}
\begin{split}
&{\beta_{4}}_4= \frac{1}{768} p \left(353792 p^{10}+6132896 p^9+42684328 p^8+155324092 p^7+322823634 p^6+396411015 p^5+\right.\\
   &\left.+294642637 p^4+138471718 p^3+41087864 p^2+7082999
   p+540225\right) \Gamma \left(2+\frac{11}{p}\right)
   \end{split}
\end{equation*}
\begin{equation*}
\begin{split}
&{\beta_{4}}_5= \frac{1}{3072} p \left(22368256 p^{12}+504188672 p^{11}+4716924384 p^{10}+24087207768 p^9+74253020268 p^8+\right.\\
   &\left.+144605869746 p^7+182135591147 p^6+151241944594 p^5+84569662551
   p^4+31866530982 p^3+\right.\\
   &\left.+7805855169 p^2+1124863038 p+72335025\right) \Gamma \left(2+\frac{13}{p}\right)
   \end{split}
\end{equation*}
\begin{itemize}
\item Series for ${\beta_{2}}_n$.
\end{itemize}
\begin{equation*}
 {\beta_{2}}_0=   \frac{1}{48} p \left(6 p^2+9 p+27\right) \Gamma \left(2+\frac{3}{p}\right)
\end{equation*}
\begin{equation*}
{\beta_{2}}_1=\frac{1}{16} p \left(16 p^4+70 p^3+111 p^2+90 p+17\right) \Gamma \left(2+\frac{5}{p}\right)
\end{equation*}
We do not continue this particular list of polynomials because for any $n$ we can use
\begin{equation*}
     {\beta_{2}}_n= \frac{{\beta_{1}}_n}{12}+\frac{ {\beta_{4}}_n}{2},
\end{equation*}
using the fact that $\beta$-contribution should vanish for Gauss-Bonnet-like combination (\ref{GB}) in the original model (see Section~\ref{SecGB}).

\begin{itemize}
\item Series for ${\gamma_{1}}_n$
\end{itemize}
\begin{equation*}
{\gamma_{1}}_1=\left(\frac{3}{2}\right)^{-\frac{6}{p}-5} p^2 \left(13 p^3+11 p^2+105 p+81\right) \Gamma \left(3+\frac{6}{p}\right)
\end{equation*}
\begin{equation*}
\begin{split}
 &{\gamma_{1}}_2= \frac{1}{9(p+4)}\left(\frac{3}{2}\right)^{-\frac{8}{p}-6} p^2 \left(873 p^6+6561 p^5+18830 p^4+38808 p^3+58775 p^2+24939
   p+\right.\\
   &\left.+1694\right)\Gamma \left(3+\frac{8}{p}\right)
   \end{split}
\end{equation*}
\begin{equation*}
 \begin{split}
&{\gamma_{1}}_3= \frac{1}{3^4} \left(\frac{3}{2}\right)^{-\frac{10}{p}-5} p^2 \left(27063 p^7+164277 p^6+333536
   p^5+411514 p^4+393359 p^3-176615 p^2-\right.\\
   &\left.-99406 p-5408\right) \Gamma
   \left(3+\frac{10}{p}\right)
   \end{split}
\end{equation*}
\begin{equation*}
 \begin{split}
&{\gamma_{1}}_4= \frac{1}{9(p+6)} \left(\frac{3}{2}\right)^{-\frac{12}{p}-7} p^3 \left(24159 p^9+251941 p^8+5724 p^7-8910634 p^6-44004219
   p^5+\right.\\
   &\left.-100165311 p^4-133974594 p^3-107395632 p^2-34345134 p-3659580\right) \Gamma \left(3+\frac{12}{p}\right)
   \end{split}
\end{equation*}
\begin{equation*}
 \begin{split}
&{\gamma_{1}}_5= \frac{1}{3^9} \left(\frac{3}{2}\right)^{-\frac{14}{p}-5} p^2 \left(-406501821 p^{11}-9510519123 p^{10}-88989177195 p^9-439309907295
   p^8-\right.\\
   &\left.-1272350395453 p^7-2285301835469 p^6-2598701292105 p^5-1735969075869 p^4-472784264634 p^3-\right.\\
   &\left.-1766424996
   p^2+19123297960 p+2201024000\right) \Gamma \left(3+\frac{14}{p}\right)
   \end{split}
\end{equation*}
\begin{itemize}
\item Series for ${\gamma_{2}}_n$
\end{itemize}
\begin{equation*}
{\gamma_{2}}_1=\frac{1}{9}\left(\frac{3}{2}\right)^{-\frac{6}{p}-3} p^2 \left(13 p^3+35 p^2+45 p+45\right)
   \Gamma \left(3+\frac{6}{p}\right)
\end{equation*}
\begin{equation*}
\begin{split}
&{\gamma_{2}}_2=\frac{1}{81 (p+4)}\left(\frac{3}{2}\right)^{-\frac{8}{p}-4} p^2 \left(873 p^6+8073 p^5+27542 p^4+47004 p^3+46379 p^2+21423
   p+\right.\\
   &\left.+2282\right) \Gamma \left(3+\frac{8}{p}\right)
   \end{split}
\end{equation*}
\begin{equation*}
\begin{split}
&{\gamma_{2}}_3=\frac{1}{3^6}\left(\frac{3}{2}\right)^{-\frac{10}{p}-3} p^2 \left(27063 p^7+216765 p^6+650516 p^5+1019878 p^4+1051859 p^3+\right.\\
   &\left.+669805
   p^2+199658 p+17176\right) \Gamma \left(3+\frac{10}{p}\right)
   \end{split}
\end{equation*}
\begin{equation*}
\begin{split}
&{\gamma_{2}}_4=\frac{1}{3^4(p+6)}\left(\frac{3}{2}\right)^{-\frac{12}{p}-5} p^2 \left(24159 p^{10}+344677 p^9+1544568 p^8+1065326 p^7-10009335
   p^6-\right.\\
   &\left.-27705015 p^5-25616142 p^4-6132024 p^3+4185918 p^2+2971404 p+629856\right) \Gamma \left(3+\frac{12}{p}\right)
   \end{split}
\end{equation*}
\begin{equation*}
\begin{split}
&{\gamma_{2}}_5=\frac{1}{3^{11}}\left(\frac{3}{2}\right)^{-\frac{14}{p}-3} p^2 \left(-406501821 p^{11}-8882686827 p^{10}-79225353279 p^9-374539543467
   p^8-\right.\\
   &\left.-1020700233661 p^7--1638034515581 p^6-1527090090477 p^5-788646121977 p^4-\right.\\
   &\left.-188566254498 p^3-978484788 p^2+7288115080
   p+607712000\right) \Gamma \left(3+\frac{14}{p}\right)
   \end{split}
\end{equation*}
\begin{itemize}
\item Series for ${\gamma_{4}}_n$
\end{itemize}
\begin{equation*}
{\gamma_{4}}_1=\left(\frac{3}{2}\right)^{-\frac{6}{p}-6} p \left(13 p^4+164 p^3+672 p^2+1008 p+189\right) \Gamma
   \left(1+\frac{6}{p}\right)
\end{equation*}
\begin{equation*}
\begin{split}
&{\gamma_{4}}_2=\frac{1}{9}\left(\frac{3}{2}\right)^{-\frac{8}{p}-7} \left(873 p^7+15813 p^6+102854 p^5+303748 p^4+384683 p^3+194395 p^2+30458 p+\right.\\
&\left.+1708\right) \Gamma \left(1+\frac{8}{p}\right)
\end{split}
\end{equation*}
\begin{equation*}
\begin{split}
&{\gamma_{4}}_3=\frac{1}{3^{4}}\left(\frac{3}{2}\right)^{-\frac{10}{p}-6} \left(27063 p^9+648954 p^8+5826731
  p^7+25813720 p^6+60609609 p^5+72209658 p^4+\right.\\
   &\left.+42934449 p^3+14216004
   p^2+2027476 p+206240\right) \Gamma \left(1+\frac{10}{p}\right)
   \end{split}
\end{equation*}
\begin{equation*}
\begin{split}
&{\gamma_{4}}_3=\frac{1}{3^{4}}\left(\frac{3}{2}\right)^{-\frac{10}{p}-6} \left(27063 p^9+648954 p^8+5826731
  p^7+25813720 p^6+60609609 p^5+72209658 p^4+\right.\\
   &\left.+42934449 p^3+14216004
   p^2+2027476 p+206240\right) \Gamma \left(1+\frac{10}{p}\right)
   \end{split}
\end{equation*}
\begin{equation*}
\begin{split}
&{\gamma_{4}}_4=\frac{1}{9}\left(\frac{3}{2}\right)^{-\frac{12}{p}-8} \left(24159 p^{11}+680953 p^{10}+7058532 p^9+34983293 p^8+86515776
   p^7+86849955 p^6-\right.\\
   &\left.-33692652 p^5-87689385 p^4+9976365 p^3+28063584 p^2+14110524 p+2834352\right) \Gamma
   \left(1+\frac{12}{p}\right)
   \end{split}
\end{equation*}
\begin{equation*}
\begin{split}
&{\gamma_{4}}_5=\frac{1}{3^9}\left(\frac{3}{2}\right)^{-\frac{14}{p}-6} \left(-406501821 p^{13}-17105308920 p^{12}-293471153478 p^{11}-2721885160236
   p^{10}-\right.\\
   &\left.-15119232832624 p^9-52373885230232 p^8-113883595585406 p^7-151988344517548 p^6-\right.\\
   &\left.-116529530763147
   p^5-45913979420376 p^4-8385606974468 p^3+206921080256 p^2+342592970240 p+\right.\\
   &\left.+26123776000\right) \Gamma
   \left(1+\frac{14}{p}\right)
   \end{split}
\end{equation*}

\bibliographystyle{JHEP}
\bibliography{BH.bib}


\end{document}